\documentclass[prd,twocolumn,showpacs,superscriptaddress,nofootinbib]{revtex4-2}

\usepackage{amsfonts}
\usepackage{amsmath}
\usepackage{amssymb}
\usepackage{amsthm}
\usepackage{bm}
\usepackage{dcolumn}
\usepackage{epsfig}
\usepackage{graphicx}
\usepackage{graphics}
\usepackage[latin1]{inputenc}
\usepackage{latexsym}
\usepackage{rotating}
\usepackage{hyperref}
\usepackage{ulem}
\usepackage{lipsum}
\usepackage[dvipsnames, usenames]{xcolor}
\usepackage{comment}

\newcommand\be{\begin{equation}}
\newcommand\ba{\begin{eqnarray}}
\newcommand\ee{\end{equation}}
\newcommand\ea{\end{eqnarray}}
\newcommand\bal{\begin{align}}
\newcommand\eal{\end{align}}
\newcommand{\pont}{{\,^\ast\!}R\,R}

\newcommand{\EH}{{\mbox{\tiny EH}}}
\newcommand{\K}{{\mbox{\tiny K}}}
\newcommand{\ergo}{{\mbox{\tiny ergo}}}
\newcommand{\dCS}{{\mbox{\tiny dCS}}}
\newcommand{\Hor}{{\mbox{\tiny H}}}
\newcommand{\fz}{{\mbox{\tiny FZ}}}
\newcommand{\nz}{{\mbox{\tiny NZ}}}
\newcommand{\maxi}{{\mbox{\tiny max}}}
\newcommand{\p}{{\mbox{\tiny +}}}
\newcommand{\m}{{\mbox{-}}}
\newcommand{\resum}{{\mbox{\tiny resum}}}

\newcommand{\Brown}{Brown Theoretical Physics Center and Department of Physics, Brown University, Providence, Rhode Island 02903, United States.}

\begin{document}
\title{Black Hole Superradiance in Dynamical Chern-Simons Gravity}

\author{Stephon Alexander}
\affiliation{\Brown}
\affiliation{CCA, Flatiron Institute, 162 5th Ave, NY, New York, United States.}

\author{Gregory Gabadadze}
\affiliation{Center for Cosmology and Particle Physics, Department of Physics, New York University, New York, New York 10003, United States.}
\author{Leah Jenks}
\affiliation{\Brown}
\author{Nicol\'as Yunes}
\affiliation{Illinois Center for Advanced Studies of the Universe, Department of Physics, University of Illinois at Urbana-Champaign, Urbana, Illinois 61801, United States.}

\date{\today}

\preprint{}

\begin{abstract}
  Black hole superradiance provides a window into the dynamics of light scalar fields and their interactions close to a rotating black hole. Due to the rotation of the black hole, the amplitude of the scalar field becomes magnified, leading to a ``black hole bomb'' effect.  Recent work has demonstrated that rotating black holes in dynamical Chern-Simons gravity possess unique structures, the ``Chern-Simons caps," which may influence the behavior of matter near the black hole. Motivated by the presence of these caps, we study superradiance in dynamical Chern-Simons gravity in the context of a slowly rotating black hole. We find additional contributions to the superradiance beyond what is expected for a Kerr black hole. Studying the superradiant spectrum of perturbations, we find that the Chern-Simons contributions give rise to small corrections to the angular dependence of the resulting scalar cloud. Finally, we comment on potential observable consequences and future avenues for investigation.
\end{abstract}


\maketitle

\allowdisplaybreaks[4]

\section{Introduction}
\label{intro}

Dynamical Chern-Simons (dCS) gravity \cite{Jackiw:2003pm, Alexander:2009tp} is an extension of general relativity (GR) in which the parity-violating Chern-Simons form coupled to a pseudo-scalar field is added to the Einstein-Hilbert action of GR. This addition is well motivated from particle physics \cite{ALVAREZGAUME1984269} and string theory \cite{Polchinski:1998rr, PhysRevLett.96.081301, Green:1987mn, Alexander:2004xd} perspectives. While non-rotating black holes in dCS gravity possess the same properties as those in GR \cite{Grumiller:2007rv,Shiromizu:2013pna}, those which are rotating are subject to corrections to the metric that are sourced by the Chern-Simons pseudo-scalar \cite{Yunes:2009hc, Yagi:2012ya,Maselli:2017kic}. As such, dCS gravity provides a vast playground on which to study deviations from GR and has been considered in a wide variety of gravitational and astrophysical contexts \cite{Alexander:2007zg, Alexander:2007vt, Alexander:2008wi, Sopuerta:2009iy, cardosogualtieri, konnoBH, amarilla, chen, Garfinkle:2010zx, Molina:2010fb, Yagi:2013mbt, Loutrel:2018ydv, Delsate:2018ome, Wagle:2019mdq, Wagle:2021tam, Doneva:2021dcc}. The most stringent constraints on dCS gravity arise from combining gravitational wave observations and X-ray observations of isolated pulsars \cite{Silva:2020acr}, while solar system experiments and pure gravitational wave observations or binary pulsar tests generally have so far left the theory unconstrained \cite{Nakamura:2018yaw,Yagi:2013mbt,Nair:2019iur,Perkins:2021mhb}. 

DCS gravity can be reliably treated as a low-energy effective field theory (EFT) valid below a certain energy scale, $\mu$, that can be near, or below the Planck scale \cite{Alexander:2021ssr}. In the past and present work we consider $\mu$ to be below the Planck scale by at least a few orders of magnitude,   $\mu \ll M_{\rm P}$. The dCS term itself is the first leading term among 
an infinite number of terms of the EFT. All these terms can be obtained from 
a conventional high-energy `progenitor' theory of a complex scalar field coupled to fermions which are coupled to each other via a Yukawa coupling and also coupled to gravity. The quantum field theory is valid above $\mu$ and below the Planck scale (see \cite{Alexander:2021ssr}). The anomalous triangle diagram of this progenitor theory 
is what is responsible for generating the dCS term in the low energy approximation.

The above comments should  make it clear that certain classical solutions -- which  emerge in dCS theory due to the higher derivative nature of its equations of motion -- are connected to physics at the scale $\mu$, and therefore,  cannot be justified since they would be modified by the infinite 
number of  higher dimensional  terms, if the latter were kept in the theory. 
Likewise, any statement about ghost states of mass $\mu$, 
inferred from the  dCS theory would not be justified.  Moreover, the high energy 
theory that completes dCS at the scale $\mu$ does not have such ghost states in its spectrum \cite{Alexander:2021ssr}. The solutions  that can reliably be discussed within dCS theory
are the ones present in GR but now  modified  due to the dCS term treated 
as a small correction. We  will continue to be interested only 
in these solutions. One may be concerned that the size of the aforementioned corrections are too small to be relevant, such as in the case of the higher-derivative terms which arise in the low-energy limit of string theory. However, we emphasize that we are considering the low-energy limit of a different high-energy theory in which the dCS term arises. In this case, while the dCS term is still suppressed by the cutoff scale, $\mu$, we have significant freedom in choosing $\mu$ (as long as $\mu \ll M_{\rm P})$ such that the contribution can be non-negligible.

Recently, we showed that dCS gravity naturally possesses a mathematical structure that leads to the violation of the conditions for geodesic focusing. More precisely, slowly rotating black holes in the theory become endowed with two ``cap-like'' structures at the north and south poles in which the focusing theorem is violated \cite{Alexander:2021ssr}. It is thus of significant interest to investigate the behavior of matter or fields near these black hole regions. While these caps are small regions compared to the size of the black hole, if they alter the behavior of matter in a meaningful way, it may be possible to point to observable signatures that could be distinct from those in GR. 

One obvious physical process that might lead to deviations from GR is black hole superradiance. In general, superradiance describes a radiative enhancement process that applies to many areas in physics, first put forth by Dicke, who coined the term ``superradiance'' in describing coherent radiation \cite{PhysRev.93.99}. Zeldovich then showed that superradiant scattering could also be sourced by a rotating surface \cite{1971JETPL..14..180Z,1972JETP...35.1085Z}. In particular, ultralight scalar fields can exhibit superradiance when they scatter off a rotating black holes, leading to a growing instability of this field.  The latter can lead to ``clouds'' around the rotating black hole that have a variety of observable implications. 

Mathematically, the study of the superradiant growth of a massive scalar field around a rotating black hole requires the solution to the Klein-Gordon equation on a rotating black hole background. The machinery for finding these solutions was first developed by Starobinsky \cite{Starobinsky:1973aij}. Black hole superradiance has since been studied for a wide range of cases, including for scalars in a variety of limits \cite{Press:1972zz, Bardeen:1972fi, Starobinsky:1973aij, Ternov:1978gq, Zouros:1979iw, Detweiler:1980uk, Dolan:2007mj,Arvanitaki:2009fg, Arvanitaki:2010sy, Dolan:2012yt,Yoshino:2013ofa,Brito:2014wla, Brito:2015oca, Baryakhtar:2020gao}, as well as for vector and tensor fields \cite{Rosa:2011my, Pani:2012bp,Pani:2012vp,East:2017mrj,Baryakhtar:2017ngi,Baumann:2019eav,Brito:2013wya,Brito:2020lup}. Superradiance has also been studied in a range of modified theories of gravity \cite{Zhang:2020sjh,Khodadi:2021owg, Khodadi:2020cht,Zhang:2021btn, Khodadi:2021mct}, however a full analytical analysis of black hole superradiance in dCS gravity is missing from the literature until now. 

Black hole superradiance is an example of the Penrose process, in which energy and angular momentum can be extracted from a rotating black hole \cite{Penrose:1971uk}. As such, it provides a landscape to probe unseen scalar sectors. In particular, superradiance is a powerful tool that can be used to constrain ultra-light dark matter (see e.g. \cite{Hui:2021tkt} for a review). The hypothesis here is that dark matter is perhaps an ultralight scalar field that grows around rotating black holes, leading to a scalar cloud that spins black holes down as it grows. This process leads to two observational consequences, the first of which arises from the emission of monochromatic gravitational waves from the  dissipation of the scalar cloud, as discussed in \cite{Arvanitaki:2010sy,Arvanitaki:2014wva, Arvanitaki:2016fyj,Arvanitaki:2016qwi, Baryakhtar:2017ngi,Brito:2017zvb, Baumann:2018vus}. The second is the spin-down of the black hole, in which the amplification of the scalar field reduces the black hole spin and mass \cite{Stott:2018opm, Ficarra:2018rfu, Brito:2014wla}. These processes can be a key tool to investigate the existence of massive particles and one could ask whether one could use them to distinguish between a Kerr black hole and a slowly rotating dCS black hole. 

In what follows, we will study a dCS black hole that is accompanied by an ultra-light scalar field in addition to the massless dCS pseudo-scalar. As an approximation, we will be working in the ``probe'' limit, in which the backreaction of the ultra-light scalar onto the spacetime itself, as well as the backreaction onto the massless dCS pseudo-scalar, can be neglected. As long as $\mu M \ll 1$ this approximation scheme is valid, so we will restrict our analysis to the parameter space in which this inequality is satisfied. We consider a scenario in which an ultra-light scalar field is amplified by the rotation of the black hole, which itself contains a massless pseudo-scalar cloud sourced by the parity-violating Pontryagin density. Working with the same machinery developed by Detweiler \cite{Detweiler:1980uk}, we solve the Klein-Gordon equation on the dCS background through a procedure of matching asymptotic expansions. We find that additional perturbations of the ultralight massive scalar contribute to this solution as compared to the solution for a ultralight massive scalar field on a Kerr background.  The ultralight massive scalar field does indeed possess an instability that is significantly dominated by the  $\ell=1, m=1, n=0$ mode, in agreement with the GR case. Lastly, we comment on how the addition of these extra modes, while small, may impact future observable endeavors. 

The structure of the paper is as follows. In Sec.~\ref{Overview}, we review dCS gravity and the properties of the ultralight scalar field. In Sec.~\ref{Detweiler}, we review the known studies of black hole superradiance in GR, focusing particularly on the well-known solution by Detweiler \cite{Detweiler:1980uk}. We then extend this analysis to dCS gravity in Sec.~\ref{dCS} and discuss specific properties of the solution in Sec.~\ref{solution}. Finally, conclude with a discussion of implications and future work in Sec.~\ref{Discussion}. Throughout the paper, we use the following conventions. We work in four spacetime dimensions with signature (-,+,+,+). Latin letters (a,b,...,h) range over all spacetime indices with round and square brackets denoting symmetrization and antisymmetrization, respectively. We work in geometric units such that $G=1=c$, unless otherwise specified. 
\section{dCS Black Hole With an External Scalar Field}
\label{Overview}

We will be considering a slowly-rotating dCS black hole which is accompanied by an ultralight scalar field in addition to the massless pseudo-scalar field associated with dCS gravity. We emphasize that this massless pseudo-scalar field is distinct from the ultralight scalar, which we will probe for superradiant behavior. The full action is as follows:
\be 
S = S_{\EH} + S_{\dCS} + S_{\vartheta} + S_{\varphi}, 
\label{eq:full-action}
\ee 
where $S_{\EH}$ is the Einstein-Hilbert action of GR, $S_{\dCS}$ contains the Chern-Simons term, $S_\vartheta$ is the action for the dCS pseudo-scalar $\vartheta$, and $S_\varphi$ is the action of the external ultralight scalar field $\varphi$. In the following, we will review the basics of dCS gravity in vacuum, as well as rotating black holes in dCS gravity, and the ultralight scalar separately, before considering the dCS black hole and scalar together. 

\subsection{Dynamical Chern-Simons Gravity in Vacuum}
\label{ABC}
The \textit{vacuum} action of dCS gravity is given by $S_{\rm vac} = S_{\EH} + S_{\dCS} + S_{\vartheta}$. Explicitly, 
\be
\label{CSaction}
S_{\rm vac} =  \int d^4x \sqrt{-g} \left[\kappa R + \frac{\alpha}{4} \vartheta \; \pont
- \frac{1}{2} \left(\nabla_a \vartheta\right) \left(\nabla^a \vartheta\right) \right]\,,
\ee
where $\kappa R$ is the usual Einstein-Hilbert term with $\kappa = (16\pi)^{-1}$ and R the Ricci scalar \cite{Jackiw:2003pm, Alexander:2009tp}. The Chern-Simons term consists of a dynamical pseudo-scalar field, $\vartheta$, which couples to the Pontryagin density of the spacetime. The Pontryagin density is defined as 
\be
\label{pontryagindef}
\pont:={\,^\ast\!}R^a{}_b{}^{cd} R^b{}_{acd}\,,
\ee
where the Hodge dual to the Riemann tensor is 
\be
\label{Rdual}
{^\ast}R^a{}_b{}^{cd}:=\frac12 \epsilon^{cdef}R^a{}_{bef}\,,
\ee
and $\epsilon^{cdef}$ is the Levi-Civita tensor. The coupling constant, $\alpha$ can be related to the cutoff of the theory, $\mu$ via $\alpha = 1/\mu$. As long as $\mu \ll M_{\rm P}$, our EFT construction remains valid \cite{Alexander:2021ssr}. The Pontryagin density can also be written in terms of divergence of the Chern-Simons topological current,   
\be
\nabla_a K^a = \frac14 \pont , 
\label{eq:curr1}
\ee
where 
\be
K^a :=\epsilon^{abcd}\left(\Gamma^n{}_{bm}\partial_c\Gamma^m{}_{dn}+\frac23\Gamma^n{}_{bm}\Gamma^m{}_{cl}\Gamma^l{}_{dn}\right)\,,
\label{eq:curr2}   
\ee
giving rise to the name ``Chern-Simons gravity''. Varying the action, Eq. \eqref{CSaction} yields the modified \textit{vacuum} field equations, 
\begin{align}\label{eq:MetricEOM1}
G_{ab} + \frac{\alpha}{\kappa} \, C_{ab} &= \frac{1}{2 \kappa} T_{ab}\,,
\end{align}
which include a modification from the C-tensor, defined as 
\begin{align}
C^{ab} = \left(\nabla_{c} \vartheta \right) \; \epsilon^{cde(a} \nabla_{e} R^{b)}{}_{d} + \left(\nabla_{c} \nabla_{d} \vartheta \right) \; {}^{\ast}R^{d(ab)c}\,.
\label{eq:C-tensor}
\end{align}
The total energy-momentum tensor $T_{ab}$ is the sum of any matter stress-energy tensor (assumed in this subsection to be zero) and the stress-energy tensor of the pseudo-scalar $\vartheta$, which is given by 
\begin{align}
T_{ab}^{(\vartheta)} = \left(\nabla_{a} \vartheta \right) \left( \nabla_{b} \vartheta \right) - \frac{1}{2} g_{ab} \left(\nabla_{c} \vartheta \right) \left(\nabla^{c} \vartheta \right)\,.
\end{align}
The pseudo-scalar field $\vartheta$ itself obeys the following \textit{vacuum} equation of motion
\begin{align}
\square \vartheta &= -\frac{\alpha}{4 \kappa} \, \pont\,,
\label{eq:theta-evolution}
\end{align}
which can be obtained by varying the action, Eq.~\eqref{CSaction}, with respect to $\vartheta$. 

\subsection{Slowly Rotating Black Holes in dCS gravity}
\label{sec:BHsol}
For spherically symmetric spacetimes, one can show that the Pontryagin density, $*RR$ must vanish ~\cite{Grumiller:2007rv, Shiromizu:2013pna}. If this is the case, and the spacetime is assumed to be static, then the scalar, $\vartheta$ must be a constant, which implies $T_{ab}=0=C_{ab}$. Then, all dCS solutions must reduce to the static and spherically symmetric solutions of GR, and more precisely, must reduce to the Schwarzschild spacetime \cite{Grumiller:2007rv}, which is a solution of dCS gravity. However, spacetimes which lack spherical symmetry, such as the Kerr solution of GR for rotating black holes, are no longer solutions of dCS gravity, even if they are stationary; this is because the Pontryagin density is nonzero in non-spherically symmetric spacetimes, and thus, it sources a non-trivial pseudo-scalar field by Eq.~\eqref{eq:theta-evolution}. 

The first such solution was found by Yunes and Pretorius ~\cite{Yunes:2009hc}, by considering a black hole in a slow rotation expansion. This solution has since been extended to second order \cite{Yagi:2012ya} and fifth order in rotation ~\cite{Maselli:2017kic}. An extremal solution for the scalar field was found in \cite{McNees:2015srl}, and a non-perturbative  numerical solution for the metric was found in~\cite{Delsate:2018ome}. The leading order correction to the metric in the slowly rotating solution is as follows:
\begin{align}
ds^{2} \!&= \!ds^{2}_{\K} \!+\! \frac{5}{4}\zeta M \chi \frac{M^4}{r^4} \left( 1+\frac{12}{7}\frac{M}{r} + \frac{27}{10} \frac{M^2}{r^2} \right) \sin^2\!\theta dt d\phi\,,
\label{metric-linear}
\end{align}
where $M$ is the ADM mass, $\zeta$ is related to the CS coupling $\alpha$ by
\be 
\zeta = \frac{\alpha^2}{\kappa M^4},
\ee 
$\chi$ is the dimensionless spin parameter, defined in terms of the ADM angular momentum via $|S| = M a = M^2 \chi$, so that $\chi = a/M$ is dimensionless, and $ds^2_\K$ is the usual line element of the Kerr metric in Boyer-Lindquist coordinates. Explicit forms for the higher-order corrections to all terms in the metric can be found in \cite{Yagi:2012ya,Maselli:2017kic}, but for simplicity we will not write them here. The associated pseudo-scalar to leading order in rotation is 
\be
\vartheta = \frac{5}{8}\alpha\chi \frac{\cos\theta}{r^2}\left( 1 + 2 \frac{M}{r} + \frac{18}{5}\frac{M^2}{r^2}\right).
\ee 
At $\mathcal{O}(\zeta\chi^2)$, this solution has an event horizon, $r_{\p}$ and an inner apparent horizon, $r_{\m}$ given by 
\be 
r_{\pm} = r_{\pm,\K} \mp \frac{915}{28672}M\zeta \chi^2, 
\label{horizon}
\ee 
where $r_{\pm, \K}$ denotes the outer and inner Kerr horizons, respectively. The location of the ergosphere is given by 
\be 
r_{\ergo} = r_{\ergo,\K} - \frac{915}{28672}M\zeta\chi^2\left( 1 + \frac{2836}{915}\sin^2\theta\right).
\label{ergo}
\ee
For the remainder of the paper we will be considering the slowly rotating dCS metric up to $\mathcal{O}(\zeta\chi^2)$, given by 
\be
g_{\mu\nu}  = g_{\mu\nu}^\K + g_{\mu\nu}^{\dCS}\left[\mathcal{O}(\zeta\chi^2)\right], 
\label{farmetric}
\ee
where $g_{\mu\nu}^K$ is the Kerr metric in Boyer-Lindquist coordinates and $g_{\mu\nu}^{\dCS}\left[\mathcal{O}(\zeta\chi^2)\right]$ is the dCS metric  correction up to quadratic order in the rotation\cite{Yagi:2012ya}. 

This metric is a valid approximation sufficiently far away from the black hole event horizon, since otherwise the ${\cal{O}}(\zeta)$ term in Eq.~\eqref{metric-linear}, which decays as $1/r^4$, would dominate over the GR term, which decays as $1/r^3$. However, close to the horizon we must take into account that its location is slightly shifted from its location in the Kerr spacetime, as we can see in Eq.~\eqref{horizon}. This shift leads to spurious divergences of Eq.~\eqref{farmetric} at the location of the Schwarzschild and Kerr horizons, which we will address by resumming the metric in $\chi$. There are in principle an infinite number of ways in which the metric can be resummed, so long as the resummed metric: 
\begin{enumerate}
\item identically reduces to the non-resummed metric, order by order in $\chi$, when expanded in small $\chi \ll 1$, 
\item has each component remain finite everywhere outside the dCS horizon, e.g. that there are no divergences at the Schwarzschild or Kerr horizon.
\end{enumerate}

One way to perform the resummation is to consider a transformation $\Delta \rightarrow \bar{\Delta}$, where $\Delta$ is defined as in the Kerr solution as 
\be 
\Delta = (r - r_\p)(r - r_{\m}) = r^2 - 2Mr + M^2\chi^2,  
\ee 
and we find the necessary shift to be 
 \be 
 \bar{\Delta} = \Delta + \frac{915}{14336}M^2\zeta\chi^2,
 \ee 
which we have found by demanding that $\bar{\Delta}(r=r_{\p,{\dCS}}) = 0$. By taking $\Delta\rightarrow \bar{\Delta}$ and the Schwarzschild factor $f \rightarrow {\bar{\Delta}}/{r^2}$ in the metric of Eq.~\eqref{farmetric}, and expanding in $\chi \ll 1$, we find that the two counterterms that we need to add in order to retain the desired asymptotic behavior are 
\be 
\delta g_{rr} = \frac{915}{14336}\frac{M^2r^2\zeta\chi^2}{\bar{\Delta}^2}, 
\ee 
and 
\be 
\delta g_{tt} = \frac{915}{14336}\frac{M^2\zeta\chi^2}{r^2}.
\ee 
We also obtain counterterms for the $g_{t\phi}$ and $g_{\phi\phi}$ components, which are 
\begin{align}
    \delta g_{t\phi} &= -\frac{915}{14336}\frac{M^3\sin^2\theta\zeta\chi^3}{r^2}, \\
    \delta g_{\phi\phi} &= \frac{915}{14336}\frac{M^4\sin^4\theta\zeta\chi^4}{r^2}.
\end{align}
We include these terms for completeness, but note that they are $\mathcal{O}(\zeta\chi^3)$ and $\mathcal{O}(\zeta\chi^4)$ and will not contribute to our resummed metric, which we limit to $\mathcal{O}(\zeta\chi^2)$. The $g_{\theta\theta}$ term is unaffected by the resummation. Then, our full resummed metric is 
\begin{align}
g_{tt,\resum} &= g_{tt,\K}(\Delta \rightarrow \bar{\Delta}) + g_{tt,\dCS}(f\rightarrow \bar{\Delta}/r^2) + \delta g_{tt}\,,
\nonumber \\
g_{rr,\resum} &= g_{rr,\K}(\Delta \rightarrow \bar{\Delta}) + g_{rr,\dCS}(f\rightarrow \bar{\Delta}/r^2) + \delta g_{rr},
\label{gresum}
\end{align}
where $g_{tt/rr,\K}$ are the usual components of the Kerr metric in Boyer-Lindquist coordinates and $g_{tt/rr,\dCS}$ is the $\mathcal{O}(\zeta\chi^2)$ dCS correction described previously. All other metric components remain unchaged at $\mathcal{O}(\zeta\chi^2)$
\subsection{dCS gravity Coupled to an Ultralight Scalar Field}

In this paper, we will not be concerned with the vacuum field equations, but rather with those arising in the presence of an additional massive scalar field $\varphi$. The action is then as given in Eq.~\eqref{eq:full-action}, where the action of the massive scalar field is
\be
S_{\varphi} = \int d^4x\sqrt{-g}\left[-\frac{1}{2}g^{ab}\nabla_a\varphi \nabla_b\varphi - \frac{1}{2}\mu^2\varphi^2\right], 
\ee 
where the mass of the scalar is given by $\mu = m \hbar $. The energy-momentum tensor associated with the massive field is then
\be 
T_{ab}^{(\varphi)} = (\nabla_a\varphi)(\nabla_b\varphi) - \frac{1}{2}g_{ab}(\nabla_a\varphi\nabla^a\varphi + \mu^2\varphi^2),
\ee
which must be added to the pseudo-scalar stress-energy tensor $T_{ab}^{(\vartheta)}$ to compose the total energy-momentum tensor $T_{ab}$. The massive scalar field $\varphi$ obeys the Klein-Gordon equation, 
\be 
(\Box - \mu^2)\varphi =0\,,
\ee 
where $\Box = \nabla_a\nabla^a$ is the d'Alembertian operator.  

In what follows, we will neglect any backreaction of the ultra-light scalar onto both the metric and the dCS pseudo-scalar, working in the probe limit. Previous work \cite{East:2013mfa, East:2017ovw} has considered the inclusion of these backreaction effects for scalar and vector fields for near-extremal black holes in GR using numerical methods. However, we will be working in the slow rotation and $\mu M \ll 1$ limit in which these effects are negligible. The equation of motion for the scalar field then becomes
\be 
(\Box_{\rm vac} - \mu^2)\varphi = 0\,,
\ee 
where the d'Alembertian operator $\Box_{\rm vac}$ is that associated with the vacuum dCS spacetime solution discussed in the previous subsection.
\section{Superradiance in GR }
\label{Detweiler}
For completeness and to make more clear the methods we will use in Sec.~\ref{dCS}, we will first review Detweiler's solution for the superradiant behavior of an ultralight scalar field in a Kerr background \cite{Detweiler:1980uk}. We will  specifically work in the $\omega M \ll 1$ and $\mu M \ll 1$ limit in order to obtain an analytical solution and determine what parameters lead to the presence of the superradiant instability, which will be determined by the imaginary part of the frequency of the radiating modes.

We begin by considering a massive scalar field which obeys the Klein-Gordon equation, 
 \be 
 (\Box_\K - \mu^2)\varphi_\K(t,r,\theta,\phi) = 0
 \label{kg}, 
 \ee 
where $\mu$ is the field mass, and $\Box_\K$ and $\varphi_\K$ denote the d'Alembertian operator and scalar field, respectively for a Kerr black hole.  The Klein-Gordon equation on a Kerr background is well studied to be separable in the  $\omega M \ll 1$ and $\mu M\ll 1$ limit by making an ansatz
\be
\varphi_\K(t,r,\theta,\phi) = \sum_{\ell=0}^{\infty}\sum_{m=-\ell}^\ell \varphi_{\ell, m}^\K,
\label{phiK}
\ee 
where 
\be 
\varphi_{\ell,m}^\K = e^{-i\omega t}e^{im\phi}S_{\ell,m}^\K(\theta)R_{\ell,m}^\K(r). 
\label{phi0far}
\ee
We can see from this expression that if the frequency, $\omega_\K$, has a positive, imaginary contribution then this expression will exponentially increase; this is the key component for the field to possess a superradiant instability. Upon applying the ansatz and separating Eq.~\eqref{kg}, we find that the resulting differential equation for $S_{\ell,m}^\K(\theta)$ gives $S_{\ell,m}^\K(\theta) = P_\ell^m(\cos\theta)$, where $P_\ell^m(\cos\theta)$ are associated Legendre polynomials. 

In order to analytically solve the radial differential equation arising from the separation of variables, we will perform a matching of asymptotic expansions. We will consider two zones: the far zone, where $r \gg M$ and the near zone, where $r - r_{\Hor,\K} \ll \rm{max}(\ell/\omega, \ell/\mu)$. We emphasize here that our use of the label ``near zone'' is distinct from the definition of the near zone in post-Newtonian theory. With the solutions in both zones in hand, we will then match the expansions of the solutions in each zone in the opposite limit inside a buffer zone, where both expansions are simultaneously valid.

\begin{figure}[htb]
    \centering
    \includegraphics[width=\linewidth]{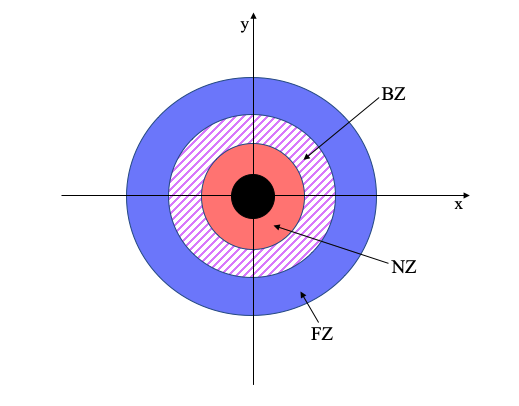}
    \caption{(Color online) Schematic diagram showing the relevant geometry. The black hole is denoted by the black circle at the center, the near zone (NZ) is in red, the far zone (FZ) is in blue and the intermediate buffer zone, where the expansions can be asymptotically matched, is shown in the purple striped region.}
    \label{zones}
\end{figure}

Figure \ref{zones} shows a schematic diagram of the three zones; this figure is meant to be an illustrative example of each of the zones, so it does not exactly represent the geometry of the black hole spacetime. The black circular region in Fig.~\ref{zones} represents the black hole. The far zone (FZ), defined as the region in which $r \gg M$, is denoted in blue, and the near zone (NZ), defined by the region in which $r - r_{\Hor,\K} \ll \mathrm{max}(\ell/\omega, \ell/\mu)$, is denoted in red. The intermediate buffer zone (BZ) is the purple striped region and is characterized by $r \gg M$ and $r - r_{\Hor,\K}\ll \mathrm{min}(\ell/\omega, \ell/\mu)$ such that near-horizon expansion of the far zone solution and the large radius expansion of the near-horizon zone solution are \textit{both} simultaneously valid and can thus be asymptotically matched.

First, we will consider the far zone, $r \gg M$. Solving the radial part of the Klein-Gordon equation in this zone and taking the appropriate boundary conditions at infinity, i.e., that the wave must be outgoing, we obtain
\be
R_{\ell}^{\K,\fz} = x^\ell e^{-x/2}U(\ell+1-\nu, 2\ell+2, x),
\label{GRSolFar}
\ee
where U is a confluent hypergeometric function \cite{AS} and we have defined 
\begin{align}
x &= 2k r \label{x},  \\
\nu &= M\mu^2/k,\label{nu}  \\
k^2 &= \mu^2 - \omega^2, \label{k} \\ 
\omega &= \sigma + i\gamma.
\label{omega}
\end{align}

Now consider the near zone, $r - r_{\Hor, \K}\ll \rm{max}(\ell/\omega, \ell/\mu)$. Let us define the quantity
\be
z_\K = \frac{r - r_{\p,\K}}{r_{\p,\K} - r_{\m,\K}},
\ee
where $r_{\p,\K}$ and $r_{\m,\K}$ are the outer and inner Kerr horizons respectively, so that the defining relation of the near zone becomes $z_\K \ll \rm{max}(\ell/\omega, \ell/\mu)$. The solution to the radial part of the Klein-Gordon equation in the near zone is then 
\be 
R_{\ell,m}^{\K,\nz} = \left(\frac{z_\K}{z_\K+1}\right)^{i P_\K} {}_2F_1(-\ell, \ell+1, \ell-2i P_\K, z_\K+1),
\label{GRSolNear}
\ee 
where again we have imposed appropriate boundary conditions at the horizon in order to obtain a single independent solution. Here, ${}_2F_1$ is the ordinary hypergeometric function and $P_\K$ is a constant, defined as 
\be
P_\K = \frac{a m - 2Mr_{\p,\K}\omega}{r_{\p,\K} - r_{\m,\K}},
\ee
where recall that $a$ is the (dimensionful) spin parameter for a Kerr black hole. 

Now, we can consider the intermediate buffer zone region. Defining the small-$x$ expansion of the far zone solution in Eq.~\eqref{GRSolFar} as $\tilde{R}_{\ell}^{\K,\fz}$, and the large-z expansion of the near-horizon solution in Eq.~\eqref{GRSolNear} as $\tilde{R}_{\ell,m}^{\K,\nz}$, asymptotic matching then requires that  
\be 
\tilde{R}_{\ell}^{\K,\fz} \sim \tilde{R}_{\ell,m}^{\K,\nz}\,,
\ee 
inside the buffer zone, where the $\sim$ symbol here means ``asymptotic to.'' This condition can only be satisfied by a set of frequencies that ensures both expressions are matched in the buffer region, and this, in turn allows us to determine the superradiance spectrum of perturbations. 

Let us now carry out this matching in detail. First, following Detweiler~\cite{Detweiler:1980uk}, we introduce the ansatz 
\be
\nu = \frac{M \mu^2}{\sqrt{\mu^2 - \omega^2}} = \ell + 1 + n + \delta\nu, 
\label{deltanu}
\ee
where the first equality comes from Eqs.~\eqref{nu}--\eqref{k}, and $\delta \nu$ is a function to be determined. With this ansatz, we use Eqs.~\eqref{nu}--\eqref{omega} to find that this function must be related to the imaginary part of the frequencies $\gamma$ via 
\be
\delta\nu = i M\gamma \left(\frac{\mu M}{l+1+n}\right)^{-3}.
\ee 
Using our knowledge that $\gamma$ scales as $(\mu M)^8$ (which we will prove in the following), it is clear that $\delta \nu = {\cal{O}}(\mu^5 M^5)$, and since $\mu M \ll 1$, then $\delta\nu \ll 1$. Assuming that ${\cal{O}}(\delta \nu) = {\cal{O}}(x^{2 \ell + 1})$ (which we will also prove in the following), we then have that the small $x$ and $\delta\nu$ expansion of Eq.~\eqref{GRSolFar} is 
\ba
\tilde{R}_{\ell}^{\K,\fz} \approx (-1)^n \frac{(2\ell+1+n)!}{(2\ell+1)!}(2k_\K r)^\ell\nonumber \\
+ (-1)^{n+1}(2\ell)!n!(2k_\K r)^{-\ell-1}\delta\nu_\K, 
\label{RFarBuffer}
\ea 
For the near-horizon solution, expanding in $z \gg 1$ gives
\begin{align} 
\tilde{R}_{\ell,m}^{\K,\nz} &\approx\frac{(-1)^\ell\Gamma(1 + 2\ell)\Gamma(1-2iP_\K)}{\Gamma(\ell+1)\Gamma(\ell+1-2iP_\K)} \left(\frac{r}{r_{\p,\K} - r_{\m,\K}}\right)^\ell \nonumber \\
&+\frac{ (-1)^{\ell-1}\Gamma(-1-2\ell)\Gamma(1-2iP_\K)}{\Gamma(-\ell)\Gamma(-\ell-2iP_\K)}\left(\frac{r}{r_{\p,\K} - r_{\m,\K}}\right)^{-\ell-1}.
\end{align}

We can now match these two asymptotic expansions to each other. That is, given two asymptotic expansions of the form $\tilde{f} \sim A \; r^l + B \; r^{-l-1}$ and $\tilde{g} \sim C \; r^l + D \; r^{-l-1}$ inside some common buffer zone, then asymptotic matching $\tilde{f} \sim \tilde{g}$ requires that $A = C$ and $B = D$. Using this to asymptotically match $\tilde{R}_{\ell}^{\K,\fz}$ and $\tilde{R}_{\ell,m}^{\K,\nz}$ we find relations for the coefficients of the $r^{\ell}$ and $r^{-\ell-1}$ terms that allow us to solve for $\delta\nu_\K$, namely
\begin{align}
 \delta \nu_\K & = (2iP_\K)\left[2k(r_{\p,\K} - r_{\m,\K})\right]^{2\ell+1} \frac{(2\ell + n +1)!}{ n!}\nonumber \\
 &\times\left[\frac{\ell!}{(2\ell+1)!(2\ell)!}\right]^2  \prod_{j=1}^l (j^2 + 4P_\K^{2}).
 \label{deltanuscaling}
\end{align}
Observe that indeed, as we expected, $\delta \nu_\K = {\cal{O}}(x^{2\ell + 1})$.

In order to find the dominant contribution to the superradiant spectrum, we need to consider the maximum positive value of $\delta\nu$. In GR, this occurs when $\ell=1$, $m=1$, and $n=0$, corresponding to 
\ba 
\gamma_\K \approx \mu \frac{(\mu M)^8}{24}\left(a/M - 2\mu r_{\p,\K}\right), 
\ea  
and a growth time of 
\ba 
\tau_\K \approx 24(\mu M)^{-8} \mu^{-1}\left(a/M - 2\mu r_{\p,\K}\right)^{-1}.
\ea 
Observe that indeed, as we expected, $\gamma_\K = {\cal{O}}[(\mu M)^8]$.
\section{ Superradiance in dCS gravity}
\label{dCS}
Having reviewed the superradiant instability for a Kerr black hole, we now turn to an investigation of superradiance in dCS gravity. We will consider the slowly rotating black hole solution in dCS gravity \cite{Yunes:2009hc,Yagi:2012ya, Maselli:2017kic} up to second order in the spin, $\mathcal{O}(\zeta\chi^2)$, appropriately resumed as explained in Sec.~\ref{sec:BHsol}. We will follow Detweiler's method \cite{Detweiler:1980uk} as described in the previous section, and determine the behavior of the scalar field by solving the Klein-Gordon equation in both the far and near zones, as defined in Fig.~\ref{zones}, but now the metric will be given by the dCS black hole solution. We will then perform asymptotic matching on the asymptotic expansions of the approximate solutions to find the relevant frequencies for the spectrum of superradiant perturbations.

\subsection{The Far Zone} 
We first consider the Klein-Gordon equation in the far zone, $r \gg M$. The metric is given in Eq.~\eqref{farmetric} because the resummed metric reduces to this equation in the far zone. With this, 
the Klein-Gordon equation becomes 
\be
(\Box_\K + \Box_{\dCS} - \mu^2)\varphi=0, 
\label{KGFar}
\ee 
to leading order in the dCS deformation, 
where $\Box_\K$ is the d'Alembertian operator of the Kerr background, $\Box_{\dCS}$ is a modification induced by the dCS corrections to the metric, and $\mu$ is again the mass of the ultralight scalar field. The full expression for $\Box_{\dCS}$ is rather complicated and can be found in Appendix \ref{BoxFull}. 

Given that the dCS contribution itself is a small correction that is subdominant to the GR contribution, we can perform an asymptotic expansion of $\Box_{\dCS}$ in the far zone and simply consider the leading order contribution. This leading order contribution, acting on a function $g(t,r,\theta,\phi) = \sum_{\ell,m} e^{-i\omega_\ell t}e^{i m\phi} g_{\ell,m}(r,\theta)$ is
\begin{align} 
\Box_{\dCS} g(t,r,\theta,\phi) &\approx  \sum_{\ell,m}\frac{603}{1792}\chi^2\zeta \frac{M^3\omega_\ell^2}{r^3}\left(\cos^2\theta - \frac{1}{3}\right)\nonumber \\
&\times g(r,\theta)e^{-i\omega_\ell t} e^{im\phi}
\label{eq:boxdCS}
\end{align}
to leading order in $r \gg M$. Observe that $\Box_{\dCS} g \sim \zeta \chi^2/r^3$, where the ${\cal{O}}(\zeta \chi)$ term does not show up because it decays as $r^{-6}$. 

In order to solve the wave equation in Eq.~\eqref{KGFar}, we will consider a perturbative solution such that
\be 
\varphi_{\fz} = \varphi_{0,\fz} + \delta\varphi_{\fz},
\label{phidecomp}
\ee 
where $\varphi_{0,\fz}$ is simply the solution on the unperturbed Kerr background and $\delta\varphi$ is the dCS correction. We take $\delta\varphi_{\fz}$ to be  $\mathcal{O}(\zeta\chi^2)$ to match the expansion order of the metric and of the d'Alembertian operator. From \cite{Detweiler:1980uk} and the discussion in Sec.~\ref{Detweiler}, we know that $\varphi_{0, \fz} $ is given by Eq. \eqref{phiK}. We can thus make a 
similar ansatz for $\delta\varphi_{\fz}$: 
\begin{align}
\delta\varphi_{\fz} &= \sum_{\ell=0}^{\infty}\sum_{m=-\ell}^\ell \left( \delta\varphi_{\ell, m}^\fz + \delta\varphi_{\ell+2,m}^\fz\right) + \sum_{\ell=2}^{\infty}\sum_{m=-\ell}^\ell \delta\varphi_{\ell-2, m}^\fz, 
\label{phifaransatz}
\end{align}
where 
\begin{align}
\delta\varphi_{\ell, m}^\fz &= e^{-i\omega_\ell t}e^{im\phi}f_{\ell,m}^\fz(r,\theta),\\
\delta\varphi_{\ell+2, m}^\fz &= e^{-i\omega_{\ell} t}e^{im\phi}f_{\ell +2,m}^\fz(r,\theta),\\
\delta\varphi_{\ell-2, m}^\fz &= e^{-i\omega_{\ell} t}e^{im\phi}f_{\ell +2,m}^\fz(r,\theta).
\label{phifar}
\end{align}

We will show soon that the solution does indeed have this structure, but one could also foresee that this must be the case due to the $\cos^2{\theta}$ dependence of Eq.~\eqref{eq:boxdCS}, which will act as a source term. This is similar to the construction of black hole perturbations for slowly rotating Kerr black holes, which leads to a coupling between $\ell$ and $\ell\pm 2$ multipoles \cite{Pani:2012bp}.
Note that in order to separate Eq.~\eqref{KGFar} in our perturbative expansion, the time dependence in $\delta\varphi_\fz$ must be the same as that of $\varphi_{0,\fz}$, namely $e^{-i\omega t}$, where the frequency $\omega$ remains the same as in the background case. In order to ensure that there is matching of the solutions and that this construction is valid, we will explicitly solve for the frequencies of the perturbed contribution via asymptotic matching and show that this is the case. 

In order to fully separate the Klein-Gordon equation in the far zone, shown in detail in Appendix \ref{LegendreRec}, we will make use of the Legendre polynomial recurrence relation to rewrite the angular dependence on the right-hand side of Eq.~\eqref{kgfar} as a linear combination of $P_\ell^m, P_{\ell + 2}^m$ and $P_{\ell-2}^m$. We thus make the following ansatzes:
\begin{align}
f_{\ell,m}^\fz(r,\theta)&=  A_{\ell,m} \, P_\ell^m(\cos\theta) g_\ell(r) 
\nonumber \\
f_{\ell -2,m}^\fz(r,\theta) &= B_{\ell,m} \, P_{\ell-2}^m(\cos\theta) g_{\ell-2}(r) 
\nonumber \\
f_{\ell+2,m}^\fz(r,\theta) &= C_{\ell,m}  \, P_{\ell+2}^m(\cos\theta)g_{\ell+2}(r),
\label{faransatz}
\end{align}
where $g_{\ell, \ell \pm 2}(r)$  describes the radial dependence and  $A_{\ell,m}, B_{\ell,m},$ and $C_{\ell,m}$ are constants given by 
\begin{align}
A_{\ell,m} &=  \frac{(\ell + \ell^2 - 3m^2)}{ 4\ell(\ell+1)-3 }, 
\nonumber \\
B_{\ell,m} &=  \frac{3(\ell-1 + m)(\ell+m)}{8\ell^2 - 2}, 
\nonumber \\
C_{\ell,m} &=  \frac{3(\ell+1-m)(2+\ell-m)}{8\ell(2 +\ell) + 6}\,.
\end{align}
These constants can be derived from the recurrence relation, shown in detail in Appendix \ref{LegendreRec}. Then, the Klein-Gordon equation separates into three separate equations, proportional to $P_\ell^m, P_{\ell + 2}^m$ and $P_{\ell-2}^m$. We will only explicitly show the solution proportional to $P_\ell^m$ mode, as the others trivially follow by substituting $\ell\rightarrow l + 2$ and $A_{\ell,m} \to B_{\ell,m}$ or $\ell\rightarrow l - 2$ and $A_{\ell,m} \to C_{\ell,m}$.

The radial equation for the $\ell$ contribution becomes
\begin{align} 
\frac{d^2}{dr^2}[r g_\ell(r)] &+ \left[\omega^2 - \mu^2 + \frac{2M\mu^2}{r} - \frac{\ell(\ell+1)}{r^2}\right]r g_\ell(r) \nonumber \\
&= -\frac{201}{1792}\frac{M^3\omega^2\zeta\chi^2}{r^2}
R_{\K, \ell}^{\fz}(r)\,.
\end{align} 
We will again define $x$ and $\nu$ as in Eqs.~\eqref{x} and \eqref{nu}
and for convenience, let us write the confluent hypergeometric function $U$ in terms of a WhittakerW function $\mathrm{W}$. Doing so, the radial equation for the $\ell$ mode becomes
\begin{align}
\frac{d^2[x g_\ell(x)]}{dx^2} &+ \left[-\frac{1}{4} + \frac{\nu}{x} - \frac{\ell(\ell+1)}{x^2}\right]xg_\ell(x)
\nonumber\\
&= -\frac{402}{1792} \frac{ M^3\omega^2 k \zeta\chi^2}{x^2}\frac{\rm{W}(\nu, \ell+1/2, x)}{x}\,,
\nonumber\\
&= -  \zeta\chi^2\Omega_1^\fz  \frac{\rm{W}(\nu, \ell+1/2, x)}{x^3}\,,
\label{radialfarde}
\end{align} 
where we have defined the overall prefactor $\Omega_1^\fz  = 402 M^3\omega^2 k/1792$.

The solution to this differential equation is the sum of the homogeneous solution and a particular solution, the later of which we shall call $g_{p, \ell}(r)$. The homogeneous solution is the same as in GR, and thus, we drop it so as to not double-count the background solution.
The particular solution is the dCS correction to the GR solution, and thus, solving Eq.~\eqref{radialfarde} with a Green functions method, we find 
\begin{align}
g_{\ell}^p &= \frac{\Omega_1^\fz \zeta\chi^2}{x}\Big[\rm{W}(\nu, \ell+1/2, x)\int \mathcal{I}_1 dx \nonumber \\
&- \rm{M}(\nu, \ell+1/2, x)\int \mathcal{I}_2 dx\Big],
\label{farsoldCS}
\end{align} 
where $\rm{W}$ and $\rm{M}$ refer to the WhittakerW and WhittakerM functions, respectively, and the integrands, $\mathcal{I}_1$ and $\mathcal{I}_2$ are complicated expressions of WhittakerM and WhittakerW functions, which can be found in detail in Appendix \ref{LegendreRec}. 

Equation~\eqref{farsoldCS} describes the radial behavior of $\varphi$ in the far zone for the $P_\ell^m$ proportional contribution. Analogous expressions can be found for the terms proportional to $P_{\ell \pm 2}^m$ by taking $\ell\rightarrow l + 2$ and $A_{\ell,m} \to B_{\ell,m}$ or $\ell\rightarrow l - 2$ and $A_{\ell,m} \to C_{\ell,m}$.
\subsection{The Near Zone}

Let us now investigate the dynamics of the scalar field in the near zone region. We will again generally follow the same method as in \cite{Detweiler:1980uk}, however we now must use the resummed metric described in Sec.~\ref{Overview} to account for the shift in the horizon location of the dCS black hole and avoid spurious divergences. Given the use of the resummed metric, we now also must account for the fact that the background solution is no longer strictly Kerr, but is the leading order contribution from the full resummed metric. 

With this in mind, we investigate the near zone ansatz
\be 
\varphi_{\nz} = \varphi_{0, \nz} + \delta\varphi_{\nz}.
\ee 
where $\varphi_{0, \nz}$ is the background solution and $\delta\varphi_{\nz}$ is a perturbation. In order to find the background solution, we will consider the Klein-Gordon equation 
\be 
\left\{\Box_{\resum}\left[\mathcal{O}(\zeta^0)\right] - \mu^2\right\} \varphi_{0,\nz} =0\,, 
\ee 
where we use the decomposition
\be
\varphi_{0,\nz} = e^{-i\omega t}e^{im\phi}S_{\ell,m}(\theta)G_{\ell,m}^{0, \nz}\,, 
\ee
for some background angular functions $S_{\ell,m}$ and some background radial functions $G_{\ell,m}^{0, \nz}$. The background Klein-Gordon equation can be separated into 
\begin{widetext}
\begin{align}
\frac{1}{\sin\theta}\frac{d}{d\theta}\left(\sin\theta \frac{dS_{\ell,m}(\theta)}{d\theta}\right) &+ \left(M^2\chi^2(\omega^2 - \mu^2)\cos^2\theta - \frac{m^2}{\sin^2\theta} + \lambda\right)S_{\ell,m}(\theta) = 0, \nonumber\\
\bar{\Delta}\frac{d}{dr}\left( \frac{\bar{\Delta}dG_{\ell,m}^{0, \nz}(r)}{dr}\right)  &+ \Big[\omega^2(r^2 + M^2\chi^2) + M^2\chi^2m^2 - \bar{\Delta}(\mu^2 r^2 + M^2\chi^2 \omega^2 + \lambda)\nonumber \\
&- 2M\chi m \omega (r^2 + M^2\chi^2 -\bar{\Delta})\Big]G_{\ell,m}^{0, \nz}(r) = 0.
\end{align}
\end{widetext}
Then, working in the $M\omega \ll 1$ and $M\mu \ll 1$ limit once again gives Legendre polynomials for the angular functions. For the radial functions, we first define $z$ such that  
\be
z = \frac{r - r_{\p}}{r_{\p} - r_{\m}},
\ee 
where $r_\p$ and $r_\m$ are the locations of the outer and inner horizons in the dCS solution, explicitly given in Eq.~\eqref{horizon}. Then, substituting for z and taking the appropriate limits, $\mu M \ll 1$, $\omega M \ll 1$, and $r-r_\p \sim z \ll max( \ell/\omega, l/\mu)$, and near-horizon boundary conditions, we find that the $\mathcal{O}(\zeta^0)$ radial solution is 
\be 
G_{\ell,m}^{0,\nz} = \left(\frac{z}{z+1}\right)^{i P} {}_2F_1(-\ell, \ell+1, \ell-2i P, z+1),
\label{O0SolNear}
\ee 
where P is now defined with the dCS horizon as
\be 
P= \frac{a m - 2Mr_{\p}\omega}{r_\p - r_\m},
\ee
and ${}_2F_1$ is again the ordinary hypergeometric function. Note that this solution has the same form as Eq.~\eqref{GRSolNear}, but $z$ and $P$ are now defined with the horizon locations of the dCS black hole rather than the Kerr solution, so it is slightly shifted from the near zone GR solution. 

Now, with the background solution in hand we can consider the full perturbative solution at $\mathcal{O}(\zeta\chi^2)$. We will consider the wave equation in the near zone using the resummed metric, Eq.~\eqref{gresum}, for which the d'Alembertian operator separates as

\be 
\Box= \Box_{\resum}\left[\mathcal{O}(\zeta^0)\right]+\Box_{\dCS}\left[\mathcal{O}(\zeta\chi^2)\right],
\ee 
where for any function $f(t,r,\theta,\phi)$ we have
\begin{widetext}
\begin{align} 
\Box_{\dCS}f(t,r,\theta,\phi) &\approx -\frac{27\omega^2M^2\zeta\chi^2}{r_{\p}^{6} \bar{\Delta}^2}\left(M - \frac{r_{\p}}{2}\right)
 \Big[M \hat{A}(r_{\p}) \cos^2\theta + \hat{B}(r_{\p})\Big]f(r,\theta)  \nonumber \\
&-\frac{\omega^2 M^2\chi^2\zeta}{150528 \bar{\Delta}^2r_\p^7}\Big[\hat{C}(r_{\p})\cos^2\theta + \hat{D}(r_{\p})\Big]f(r,\theta)\left(\frac{r - r_{\p}}{r_{\p} - r_{\m}}\right)(r_{\p} - r_{\m}).
\label{boxresum}
\end{align}
\end{widetext}
The coefficients, $\hat{A}, \hat{B}, \hat{C},$ and $\hat{D}$ are functions of $M$ and $r_\p$ and can be found in Appendix \ref{LegendreRec}.
    
Let us now consider the ansatz for the perturbation
\begin{align}
\delta\varphi_{\nz} &= \sum_{\ell=0}^{\infty}\sum_{m=-\ell}^\ell \left( \delta\varphi_{\ell,m}^\nz  +\delta\varphi_{\ell+2, m}^\nz\right) +  \sum_{\ell=2}^{\infty}\sum_{m=-\ell}^\ell \delta\varphi_{\ell-2,m}^\nz,
\end{align}
with 
\begin{align}
\delta\varphi_{\ell,m}^\nz &= e^{-i\omega_\ell t}e^{im\phi}f_{\ell,m}^\nz(r,\theta), \\
\delta\varphi_{\ell+2,m}^\nz &= e^{-i\omega_{\ell} t}e^{im\phi}f_{\ell+2,m}^\nz(r,\theta),\\
\delta\varphi_{\ell-2,m}^\nz &= e^{-i\omega_{\ell} t}e^{im\phi}f_{\ell-2,m}^\nz(r,\theta).
\label{phinear}
\end{align}

Then, as before, we can expand out the full Klein-Gordon equation, shown in Appendix \ref{LegendreRec}. Then, using our intuition about the angular structure of the dCS contribution, we can make the ansatzes
\begin{align}
f_{\ell,m}^\nz(r,\theta) &= h_{\ell,m}(r) P_\ell^m(\cos\theta)\nonumber \\
f_{\ell-2,m}^\nz(r,\theta) &=  h_{\ell-2,m}(r) P_{\ell-2}^m(\cos\theta) \nonumber \\
f_{\ell+2,m}^\nz(r,\theta) &=  h_{\ell+2,m}(r) P_{\ell+2}^m(\cos\theta), 
\end{align} 
which allows us to separate the equation.  More details can be found in Appendix \ref{LegendreRec}.

The functions $h_{\ell,m}$ and $h_{\ell \pm 2,m}$ describe the radial behavior of the perturbation. As before, let us focus on the $h_{\ell,m}$ solution, as the $h_{\ell \pm 2}$ solutions can be found by taking $\ell \rightarrow \ell \pm 2$ and the appropriate coefficients. Decoupling the three equations and writing everything in terms of z, we obtain 
\begin{widetext}
\begin{align} 
z(z+1)\frac{d}{dz}\left[z(z+1)\frac{dh_{\ell,m}(z)}{dz}\right]
&+ \left[P^{2} - \ell(\ell+1)z(z+1)\right]h_{\ell,m}(z)\nonumber \\
&= \Omega_1^\nz \tilde{A}_{\ell,m} + \Omega_2^\nz \bar{A}_{\ell,m} z)\zeta\chi^2 \frac{[z(r_{\p} - r_{\m}) + r_{\p}]^2}{ z(z+1)}\left(\frac{z}{z+1}\right)^{iP}
{}_2F_1([-\ell, \ell+1], [1 + 2iP], -z), 
\label{neardedCS}
\end{align} 
\end{widetext}
where the prefactors have been defined as $\Omega_1^\nz = 27 \omega^2 M^2/[r_{\p}^{6}(r\p - r_\m)^4]$, $\Omega_2^\nz  =\omega^2M^2/(150528 r_{\p}^{7}(r_\p - r_\m)^4 )$, and $\tilde{A}$ and $\bar{A}$ are constant coefficients which are derived from the Legendre polynomial recurrence relation given by 
\begin{align}
\tilde{A}_{\ell,m} &= \hat{B}(r_{\p}) + \hat{A}(r_{\p})\frac{(-\ell+2\ell(1+\ell) - 2m^2)}{-3 + 4\ell(\ell+1)},\\
\bar{A}_{\ell,m} &= \hat{D}(r_{\p}) + \hat{C}(r_{\p})\frac{(-\ell+2\ell(1+\ell) - 2m^2)}{-3 + 4\ell(\ell+1)},
\end{align}
whose derivations can be found in Appendix \ref{LegendreRec} along with the relevant coefficients for the $\ell \pm 2$ contributions. Solving Eq.~\eqref{neardedCS} again gives a homogeneous solution, which is just the background solution at $\mathcal{O}(\zeta^0)$, Eq.~\eqref{O0SolNear}, and we thus discard to avoid double counting. The particular solution at $\mathcal{O}(\zeta\chi^2)$, $h_{p, \ell}$, is 
\begin{widetext}
\begin{align}
h_{\ell,m}^p(z) &=
\zeta\chi^2(4P^{2} + 1)\left(\frac{z}{z+1}\right)^{iP}  {}_2F_1([-\ell, \ell+1], [1+ 2iP], -z) \int \mathcal{J}_1 dz\nonumber \\
&-\zeta\chi^2(4P^{2}+1)\Big[z(z+1)\Big]^{-iP} {}_2F_1([-\ell-2iP, \ell+1 - 2iP], [1-2iP], -z)\int \mathcal{J}_2 dz,
\label{soldCSNear}
\end{align}
\end{widetext}
where $\mathcal{J}_1$ and $\mathcal{J}_2$ are complicated expressions of hypergeometric functions, which can be found in detail in Appendix \ref{LegendreRec}. We have thus obtained the radial behavior of solution for the scalar field in the near zone, Eq.~\eqref{soldCSNear}.

\subsection{Asymptotic Matching}

With the near and far zone solutions, given by Eqs.~\eqref{soldCSNear} and \ref{farsoldCS}, in hand, we now move on to matching them asymptotically in the buffer zone. In general, both of these solutions are complicated integral expressions. In order to perform these integrals and find the conditions on the frequency to obtain the superradiant instability, we focus on an intermediate regime of overlap, the `buffer zone,' as described previously in Figure \ref{zones}. As long as $r \gg M$ and $r - r_{\Hor, \dCS} \ll \mathrm{min}(\ell/\omega, \ell/\mu)$,  both the large-z expansion of Eq.~\eqref{soldCSNear} and the small-x expansion of Eq.~\eqref{farsoldCS} will be valid. As long as $\omega M \ll 1$ and $\mu M \ll 1$, there exists a region where this is the case and the two solutions will be asymptotic to each other. 

First, let us consider the two asymptotic expansions of the approximate solution, beginning with the far zone solution. Defining $\delta\nu_\ell$ in analogy to Eq.~\eqref{deltanu}, expanding Eq.~\eqref{farsoldCS} in small $x$ and small $\delta\nu_\ell$, recalling that $\delta\nu_\ell \ll 1$, and performing the integral, we obtain 
\begin{align}
\tilde{g}_{\ell}^p(x) &\approx \frac{(1+\ell)\Gamma(-2\ell-2)}{\ell\Gamma(-1-2\ell -n) } x^{\ell -1}\Omega_1^\fz \zeta\chi^2\nonumber \\
&+ \frac{(-1)^{1+n} \ell n! \Gamma[2\ell]}{1 + \ell} \delta\nu_\ell x^{-\ell-2}\Omega_1^\fz \zeta\chi^2
\label{dCSFarSmallx}
\end{align}
For details of this expansion, see Appendix \ref{matchingdetails}. Let us now focus on the near zone solution. Similarly simplifying and expanding Eq.~\eqref{soldCSNear} in large z, we can perform the integration to obtain 

\begin{align}
    &\tilde{h}_{\ell,m}^p(z) \approx \Bigg[\frac{-2^{2\ell-1} (r_\p - r_\m)^2 \Gamma(\ell + 1/2)\Gamma(1 + 2iP)}{\ell (1 + 4 P^2) \sqrt{\pi}\Gamma(1 + l + 2iP)} z^{\ell-1}\nonumber \\
    &+ \frac{(r_\p - r_\m)^2 \Gamma(-2\ell-1)\Gamma(\ell+1)\Gamma(1 + 2iP)\sin(\ell \pi)}{(1 + 4P^2)\pi \Gamma(-\ell + 2i P)} z^{-\ell-2}\Bigg]\nonumber\\
    \times &\Omega_2^\nz \bar{A}_{\ell,m}\zeta\chi^2
    \label{dCSNearLargez}
\end{align}
The details of this expansion can also be found in Appendix \ref{matchingdetails}. Both asymptotic expansions of the approximate solutions are suppressed by an overall factor of $r$ relative to the asymptotic expansions of the GR solutions, in addition to the suppression due to the expansion in small $\zeta\chi^2$. 

Asymptotic matching requires that in the buffer zone $\tilde{g}(r) \sim \tilde{h}(r)$. Using Eqs.~\eqref{dCSFarSmallx} and~\eqref{dCSNearLargez}, we can then match coefficients to find $\delta\nu_\ell$, namely 
\begin{align}
 \delta \nu_\ell &= 2iP[2k(r_{\p} - r_{\m})]^{2\ell+1} \frac{(2\ell + n +1)!}{ n!}\nonumber \\
 &\times \left[\frac{\ell!}{(2\ell+1)!(2\ell)!}\right]^2  \prod_{j=1}^\ell (j^2 + 4P^{2}), 
 \label{deltanufinal}
\end{align}
where we have made use of the well-known identity $\Gamma(1 + x) = x\Gamma(x)$ and variations thereof. For details of this matching, see Appendix \ref{matchingdetails}. 

As expected, this expression yields the same result as for the background solution. The $\ell$th contributions are maximized for $\ell=1, m=1$ and $n=0$ and have a dominant imaginary frequency contribution and growth rate of 
\begin{align}
\gamma_\ell^\maxi &\approx \mu \frac{(\mu M)^8}{24}\left(\chi - 2\mu r_{\p}\right) \nonumber \\
\tau_\ell^\maxi &\approx 24(\mu M)^{-8} \mu^{-1}\left(\chi - 2\mu r_{\p}\right)^{-1}.
\end{align}

\section{Properties of the Superradiant Instability in dCS Gravity}
\label{solution}
Let us now consider the specific ways in which the solution found in Sec.~\ref{dCS} differs from the standard GR solution found in \cite{Detweiler:1980uk} and discussed in Sec.~\ref{Detweiler}. To recap, in general the solution can be written as
\be
\varphi = \varphi_0 + \delta \varphi = \sum_{\ell=0}\sum_{m=-\ell}^{\ell} \varphi_{\ell,m}\,,
\label{eq:solrecap}
\ee
where 
\begin{align} 
\varphi_{\ell,m} &= \Big[\varphi^0_{\ell,m}(r,\theta) + \delta\varphi_{\ell,m}(r,\theta) 
+ 
\delta\varphi_{\ell+2,m}(r,\theta) \nonumber \\
&+\delta\varphi_{\ell-2,m}(r,\theta)\Big] e^{-i \omega_{\ell} t + im\phi}.
\end{align}
Now we may consider the effects of this solution. The first question one may look to answer is which of these terms dominates the growth of the ultralight scalar. The larger $\ell$ is, the smaller the frequency $\omega_{\ell}$, and thus, the slower the mode will grow. Therefore, the dominant contributions come from the most rapidly growing modes, which have the largest frequencies and thus the smallest $\ell$s. The fastest growing mode, as in GR, is $\ell=1$. Below in Figure 2, we show the comparison between the $\ell =1,$ $\ell=2$ and $\ell=3$ frequencies. From the figure, we see that the $\ell=1$ frequency is indeed dominant, and thus the higher-$\ell$ contributions will largely be suppressed. 
 \begin{figure*}[htb]
    \centering
    \includegraphics[width=.45\linewidth]{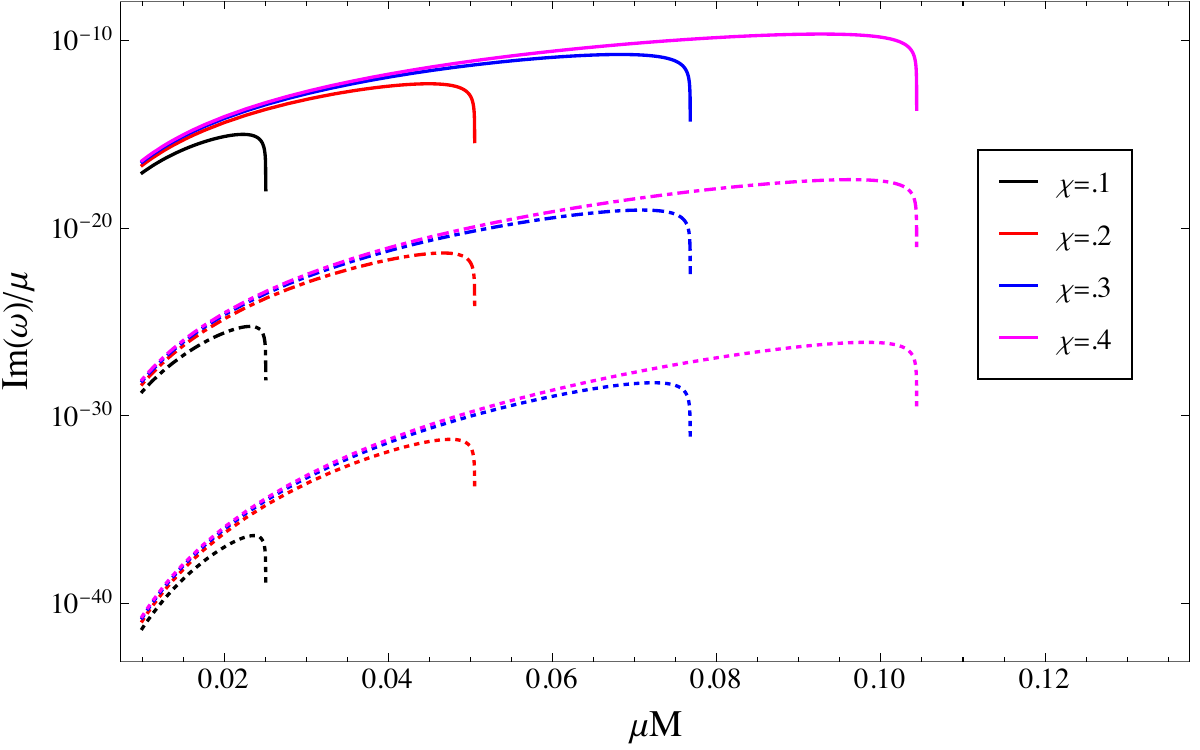}
    \caption{(Color online) Imaginary part of the frequency, Im($\omega$), for $\ell=1$ (solid), $\ell=2$ (dot-dashed), and $\ell=3$ (dotted). }
    \label{fig:wl1l3dCS}
\end{figure*}

\begin{figure*}[htb]
    \centering
    \includegraphics[width=.45\linewidth]{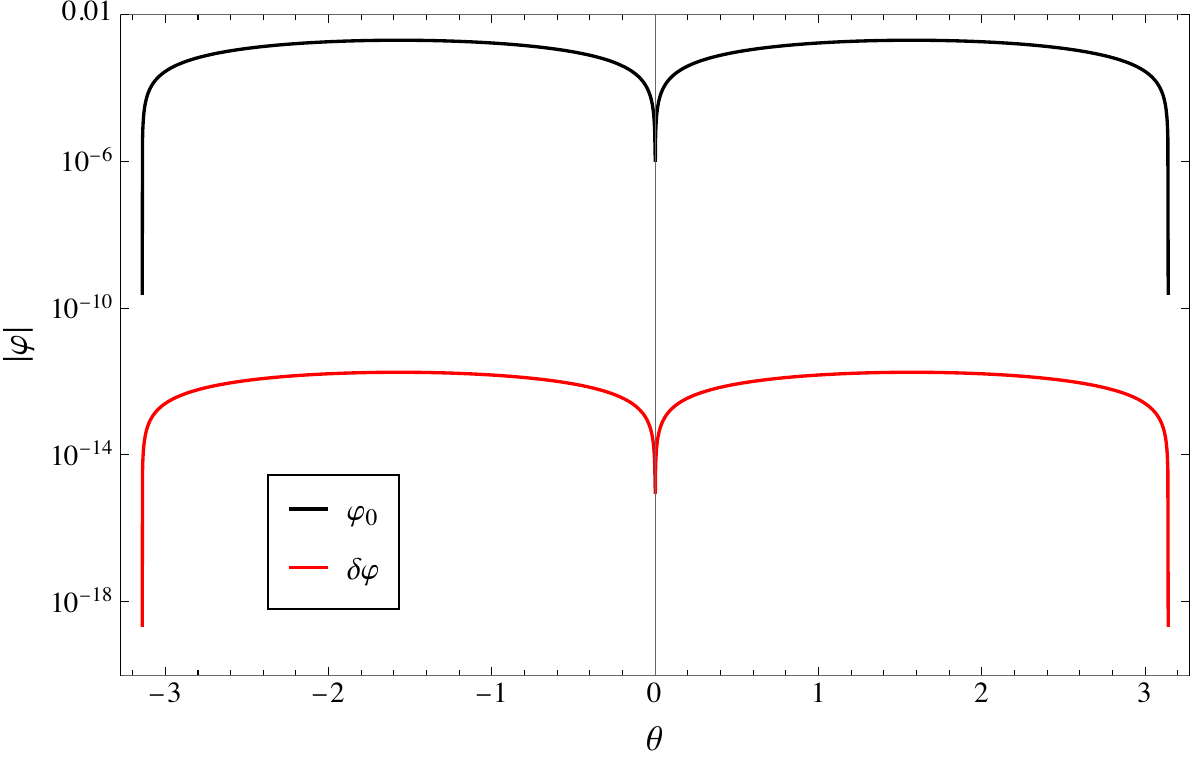}
     \includegraphics[width=.46\linewidth]{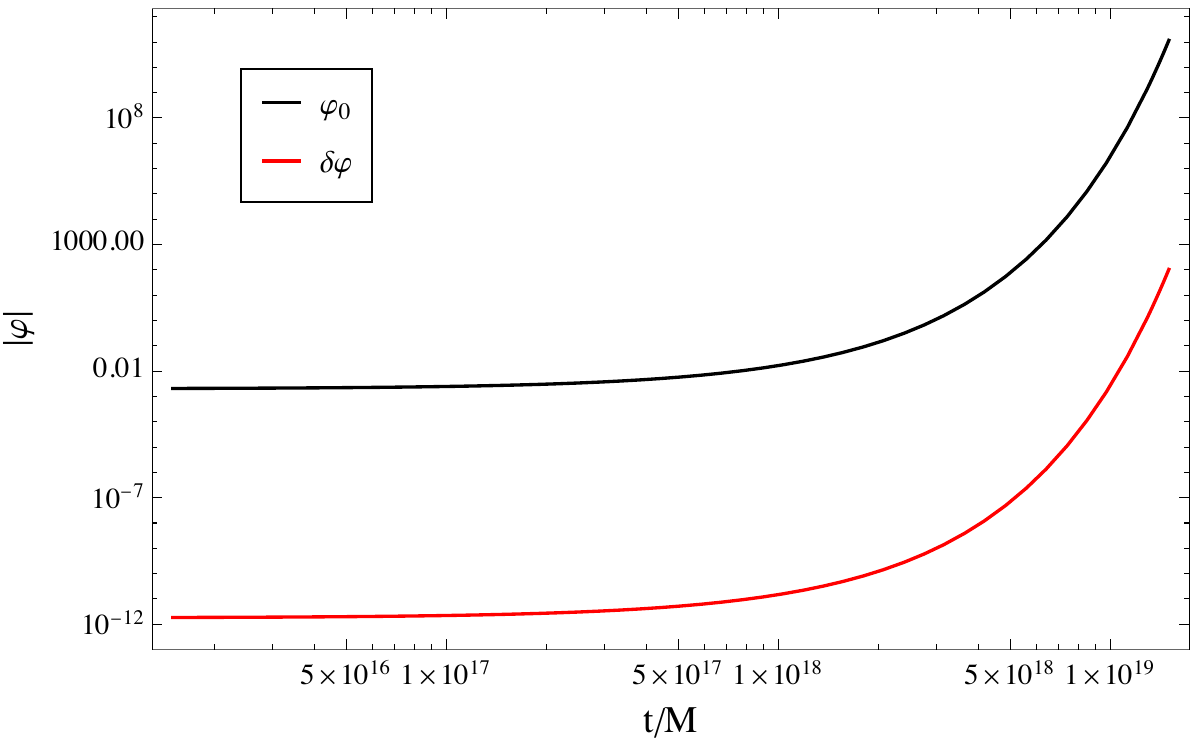}
    \caption{(Color online) Angular dependence (left) and time evolution (right) of the magnitude of the solution for $|\varphi_0|$ (black) compared to $|\delta\varphi|$ (red), where $\varphi_0$ is given by Eq.~\eqref{phi0far} and $\delta\varphi$ is given by Eq.~\eqref{RFarBuffer}. For the radial dependence, we consider the buffer zone expansion of the far zone solution, Eq.~\eqref{phifaransatz} for $\varphi_0$ and Eq.~\eqref{dCSFarSmallx} for $\delta\varphi$.  We show the solutions for the dominant mode $\ell=1$. We take $\zeta=.1, \chi=.1, M=1, \mu M= .02$, and  $r = 5 M$. For illustrative purposes we take $t/M=1$ on the left, but note that the actual timescale of the instability will be much greater. On the right we take $\theta = \pi/2$ where the amplitude of the field is maximized.}
    \label{fig:l1l3dCS}
\end{figure*}

Another question one may wish to ask is whether the angular profile of the scalar cloud is modified in dCS gravity. Although the dCS effects are perturbative, we do expect there to be a small impact on the angular behavior of the cloud. The traditional behavior of the superradiant instability in GR leads to two clouds emanating from the equator of the black hole in analogy to the p-orbital of an electron. In Figure \ref{fig:l1l3dCS}, the left-hand panel shows the angular dependence of $|\varphi_0|$ and $|\delta \varphi|$ evaluated in the buffer zone (Eq.~\eqref{phi0far} and Eq.~\eqref{phifar} respectively) for the dominant $\ell=1$ mode 
As expected, the solution is dominated by the $\ell=1$ part of $|\varphi_0|$, leading to a cloud that is still shaped like a p-orbital of an electron. There will be corrections to this, in the form of additional lower order harmonics, but these will be subdominant. Note that the amplitude of $|\delta\varphi|$ is significantly suppressed compared to that of $|\varphi^0|$. This can be seen from considering the solution in Eq.~\eqref{dCSFarSmallx} and observing that in addition to the suppression by $\zeta\chi^2$ and the additional factor of $r^{-1}$ compared to the background solution, there is an overall factor of $\omega^2k$ in the coefficient $\Omega_1^\fz$. This factor is $\mathcal{O}(\mu^3)$. All together, this gives an overall suppression of $\mathcal{O}(10^{-9})$ for the chosen values of $\mu, \zeta$ and $\chi$, which is what we observe for the $\ell=1$ mode. 

Lastly, we consider the time evolution of the solution. Taking the same buffer zone approximation discussed above, the right panel of Figure \ref{fig:l1l3dCS} compares the growth of $\varphi_0$ to that of $\delta\varphi$ on the equator for the dominant $\ell = 1$ mode.
Observe that for $\ell=1$ the growth of the instability is dominated by $\varphi_0$, again aligned with our expectations of the behavior of the field.  Although not shown in the figure, there is a large time (about $t/M \sim 10^{22}$, corresponding to $t \sim 10^9$ years, or roughly $10\%$ of a Hubble time, for a $M = 10^6 M_\odot$ black hole) at which $\delta \varphi$ dominates over $\varphi_0$ and our perturbative expansions cease to be valid. This, however, will typically occur after the superradiant instability is quenched due to the draining of spin angular momentum from the background black hole.  

\section{Discussion and Conclusions}
\label{Discussion}

We have calculated the superradiance spectrum of perturbations for a slowly rotating black hole in dCS gravity. We have found that the spectrum is composed of two contributions. For every mode with frequency $\omega_\ell$, there is a background contribution analogous to the usual GR solution and a perturbation which is proportional to a linear combination of $P_\ell^m$ and $P_{\ell \pm 2}^m$. Similar to the solution for a Kerr black hole, the spectrum is dominated by the $\ell=1, m=1, n=0$ mode. while the higher $\ell$ modes do obtain additional contributions from the dCS perturbation, they remain subdominant. The dCS corrections are subdominant to the leading order $\ell=1$ GR contribution, and thus the angular dependence and overall shape of the scalar field cloud have small corrections which induce small deviations from GR.  

Despite the lack of macroscopic distinction of the angular dependence of the scalar cloud between dCS gravity and GR, there are still many possible avenues toward determining potential observable signatures and placing constraints on dCS gravity. In general, superradiance can be used as a probe of ultralight scalars. These probes can equivalently be used to test dCS gravity by extending the analysis done for ultralight scalar superradiance sourced by a Kerr black hole to include the dCS corrections. In particular, it would be of interest to determine the energy extraction due to the amplification of the scalar field and the resulting impact on the black hole Regge plane. It has also been suggested in \cite{Ficarra:2018rfu} that including multiple modes in the analysis of the superradiant instability can impact the evolution of the scalar cloud. Presumably, the dCS solution would have a similar effect. This in turn, could lead to stronger spontenous emission of gravitational waves, as the scalar cloud transitions between energy levels. These ideas could be interesting avenues for future exploration.

A puzzling finding of our results is that presence of the Chern-Simons caps do not seem to have an effect on the superradiant instability for the dCS black hole. One might expect that the behavior of the scalar field near the black hole would be altered due to the unique properties of the caps; however, we have found that the angular behavior of the scalar field in dCS gravity is similar to that in GR. One possible explanation is that the caps are located at the north and south poles of the black hole and extend out to about twenty degrees on each side. Superradiance is a process that is sourced by the rotation of the black hole, which is minimized at the poles where the effects of the caps are at a maximum. Due to this non-overlap of regions, the caps appear to be not relevant to the presence of the superradiant instability. 

There are other avenues for future work related to superradiance in dCS gravity. One such avenue is to consider whether fermionic superradiance arises in dCS gravity. Fermions have been extensively studied in dCS gravity, in e.g. \cite{Alexander:2008wi}. Given that fermions do not exhibit superradiant behavior in GR \cite{Unruh:1974bw}, the presence of fermionic superradiance would be an excellent test of dCS gravity from which one could potentially obtain an observable signature. One could also explore other couplings and interactions of the scalar field, such as a conformal coupling, or coupling between the Chern-Simons pseudo-scalar and the external ultralight scalar field. Lastly, now that we have determined that Chern-Simons caps do not play a role in the superradiance process in dCS, it may be of interest to explore other phenomena in which the caps impact the dynamics of matter near dCS black holes. 

\acknowledgements
We thank Helvi Witek for useful discussions. We also thank an anonymous referee for pointing out an error in the first version of this paper. GG is supported by NSF grant PHY-1915219. SA, LJ and NY are supported by Simons Foundation, Award number 896696. 

\appendix
\section{Details of Klein-Gordon Separation}
\label{LegendreRec}
Here we provide details of the procedure to separate the Klein-Gordon equation obtained in Sections \ref{dCS}. We also provide details of the Legendre polynomial reccurrence relation and the derivation of the constant coefficients in Eq.~\eqref{faransatz} and Eq.~\eqref{neardedCS}. 

\subsection{Far Zone}
With the ansatz in Eq.~\eqref{phidecomp}, the Klein-Gordon equation in the far zone reduces to 
\begin{widetext}
\begin{align}
&r^2 \Bigg[\frac{1}{\sin\theta}\frac{\partial}{\partial\theta}\left(\sin\theta \frac{\partial f_{\ell,m}^\fz(r,\theta)}{\partial\theta}\right)
-  m^2 \csc^2\theta f_{\ell,m}^\fz(r,\theta)\Bigg] + r^2 \frac{\partial}{\partial r}\left[r^2 \frac{\partial f_{\ell,m}^\fz(r,\theta)}{\partial r}\right]
+ \left(\omega_\ell^2 r^4 - \mu^2r^4 + 2M\mu r^3\right)f_{\ell,m}^\fz(r,\theta) \nonumber \\
&+r^2 \Bigg[\frac{1}{\sin\theta}\frac{\partial}{\partial\theta}\left(\sin\theta \frac{\partial f_{\ell+2,m}^\fz(r,\theta)}{\partial\theta}\right)
-  m^2 \csc^2\theta f_{\ell+2,m}^\fz(r,\theta)\Bigg] + r^2 \frac{\partial}{\partial r}\left[r^2 \frac{\partial f_{\ell+2,m}^\fz(r,\theta)}{\partial r}\right] \nonumber \\
&+ \left(\omega_{\ell}^2 r^4 - \mu^2r^4 + 2M\mu r^3\right)f_{\ell+2,m}^\fz(r,\theta) \nonumber \\
&+ r^2 \Bigg[\frac{1}{\sin\theta}\frac{\partial}{\partial\theta}\left(\sin\theta \frac{\partial f_{\ell-2,m}^\fz(r,\theta)}{\partial\theta}\right)
-  m^2 \csc^2\theta f_{\ell-2,m}^\fz(r,\theta)\Bigg] + r^2 \frac{\partial}{\partial r}\left[r^2 \frac{\partial f_{\ell-2,m}^\fz(r,\theta)}{\partial r}\right]\nonumber \\
&+ \left(\omega_{\ell}^2 r^4 - \mu^2r^4 + 2M\mu r^3\right)f_{\ell-2,m}^\fz(r,\theta) 
\nonumber \\
&= -\frac{201M^3}{1792}r\omega_\ell^2 (3\cos^2\theta -1) P_\ell^m(\cos\theta)R_{\K, \ell}^\fz (r)\zeta\chi^2,
\label{kgfar}
\end{align}
\end{widetext}
where $R_{\K,\ell}(r)$ is given by Eq.~\eqref{GRSolFar}.

In Eq.~\eqref{kgfar} on the right-hand side of our expression for the far zone, we have angular dependence $(3\cos^2\theta - 1)P^m_\ell(\cos\theta)$. Employing the recurrence relation
\begin{widetext}
\begin{align}
(2\ell+1)\cos\theta P_{\ell}^{m}(\cos\theta) &= (\ell-m+1)P_{\ell+1}^m(\cos\theta) + (\ell+m)P_{\ell-1}^m(\cos\theta)
\label{PlmRec}
\end{align} 
\end{widetext}
twice to the first term and collecting coefficients, we find that 
\begin{align}
(3\cos^2\theta - 1)&P_{\ell}^{m}(\cos\theta) = \frac{(\ell + \ell^2 - 3m^2)}{ 4\ell(\ell+1)-3 }P_\ell^m(\cos\theta) \nonumber \\
&+ \frac{3(\ell-1 + m)(\ell+m)}{8\ell^2 - 2}P_{\ell-2}^m(\cos\theta)\nonumber \\
&+ \frac{3(\ell+1-m)(2+\ell-m)}{8\ell(2 +\ell) + 6}P_{\ell+2}^m(\cos\theta).
\end{align}
We then define the coefficients
\begin{align}
A &=  \frac{(\ell + \ell^2 - 3m^2)}{ 4\ell(\ell+1)-3 }, \\
B &=  \frac{3(\ell-1 + m)(\ell+m)}{8\ell^2 - 2},
\end{align} 
and
\begin{align} 
C =  \frac{3(\ell+1-m)(2+\ell-m)}{8\ell(2 +\ell) + 6}.
\end{align}
We can then separate Eq.~\eqref{kgfar} into three radial equations. Lastly, we present the full expressions for the integrands of Eq.~\eqref{farsoldCS} as follows:
\begin{widetext}
\begin{align}
\mathcal{I}_1 &= \frac{\rm{M}(\nu, \ell+1/2, x)\rm{W}(\nu, \ell+1/2, x)}{x^2\Big[\rm{W}(\nu, \ell+1/2, x)\rm{M}(1+\nu, \ell+1/2, x)
(1 + \nu + \ell)
+ \rm{M}(\nu, \ell+1/2, x)\rm{W}(1 + \nu, \ell+1/2,x)\Big]}, \\
\mathcal{I}_2 &= \frac{\rm{W}(\nu, \ell+1/2, x)^2}{x^2\Big[\rm{W}(\nu, \ell+1/2, x)\rm{M}(1+\nu, \ell+1/2, x)
(1 + \nu + \ell)
+ \rm{M}(\nu, \ell+1/2, x)\rm{W}(1 + \nu, \ell+1/2,x)\Big]}. 
\end{align} 
\end{widetext}

\subsection{Near Zone}
First, note that the constant coefficients in the expression for $\Box_{\dCS}$ for the resummed metric are 
\begin{widetext}
\begin{align}
\hat{A}(r_{\p}) &= -{\frac {149\,{M}^{5}r_{\p}}{2016}}+{\frac {445\,{M}^{4}r_{\p}^{
2}}{2016}}+{\frac {235\,{M}^{3}r_{\p}^{3}}{1568}}+{\frac {3725\,{M
}^{2}r_{\p}^{4}}{21168}}+{\frac {67\,Mr_{\p}^{5}}{896}}+{M}^{6
}+{\frac {67\,r_{\p}^{6}}{2688}},
\\
\hat{B}(r_{\p}) &= -{\frac {{M}^{7}}{3}}-{\frac {733\,{M}^{6}r_{\p}}{6048}}-{\frac {155
\,{M}^{5}r_{\p}^{2}}{2016}}-{\frac {2785\,{M}^{4}r_{\p}^{3}}{
254016}}-{\frac {11075\,{M}^{3}r_{\p}^{4}}{508032}}-{\frac {97\,{M
}^{2}r_{\p}^{5}}{16128}}+{\frac {37\,Mr_{\p}^{6}}{32256}}+{
\frac {305\,r_{\p}^{7}}{64512}}, \\
\hat{C}(r_{\p}) &= 24385536\,{M}^{8} -
11662560\,{M}^{7} {\it 
r_\p}+4189248\, {M}^{6}{{
\it r_\p}}^{2}+481680\, {
M}^{5}{{\it r_\p}}^{3}\nonumber \\
&+821280\, {M}^{4}{{\it r_\p}}^{4}-53688\, {M}^{3}{{\it r_\p}}^{5}+50652\,M{{\it r_\p}}^{7}, \\
\hat{D}(r_{\p}) &= -8128512\,{M}^{8}+924000\,{M}^{7}{\it r_\p}-264768\,{M}^{6}{{\it r_\p}}^{2
}+335040\,{M}^{5}{{\it r_\p}}^{3}\nonumber \\
&-132640\,{M}^{4}{{\it r_\p}}^{4}+19856\,{
M}^{3}{{\it r_\p}}^{5}-16884\,M{{\it r_\p}}^{7}+19215\,{{\it r_\p}}^{8}.
\end{align}
\end{widetext}
Then, we have the full Klein-Gordon Equation,
\begin{widetext}
\begin{align}
&\frac{1}{\bar{\Delta}\Sigma}\Bigg\{\Big[(M^2\chi^2 + r^2)^2 - M^2\chi^2\bar{\Delta}\sin^2\theta\Big]\omega_\ell^2 - 2M\chi m\omega_\ell(M^2\chi^2 + r^2 - \bar{\Delta})  \nonumber \\
&- \Big(\bar{\Delta}\csc^2\theta - \chi^2M^2\Big)m^2 + \bar{\Delta}\partial_\theta^2+ \bar{\Delta}^2\partial_r^2  + \bar{\Delta}\cot\theta \partial_\theta + 2(r-M)\bar{\Delta} \partial_r\Bigg\}f_{\ell,m}^\nz(r,\theta)\nonumber \\
&+\frac{1}{\bar{\Delta}\Sigma}\Bigg\{\Big[(M^2\chi^2 + r^2)^2 - M^2\chi^2\bar{\Delta}\sin^2\theta\Big]\omega_{\ell}^2 - 2M\chi m\omega_{\ell}(M^2\chi^2 + r^2 - \bar{\Delta})  \nonumber \\
&- \Big(\bar{\Delta}\csc^2\theta - \chi^2M^2\Big)m^2 + \bar{\Delta}\partial_\theta^2+ \bar{\Delta}^2\partial_r^2  + \bar{\Delta}\cot\theta \partial_\theta + 2(r-M)\bar{\Delta} \partial_r\Bigg\}f_{\ell-2,m}^\nz(r,\theta)\nonumber\\
&+\frac{1}{\bar{\Delta}\Sigma}\Bigg\{\Big[(M^2\chi^2 + r^2)^2 - M^2\chi^2\bar{\Delta}\sin^2\theta\Big]\omega_{\ell}^2 - 2M\chi m\omega_{\ell}(M^2\chi^2 + r^2 - \bar{\Delta})  \nonumber \\
&- \Big(\bar{\Delta}\csc^2\theta - \chi^2M^2\Big)m^2 + \bar{\Delta}\partial_\theta^2+ \bar{\Delta}^2\partial_r^2  + \bar{\Delta}\cot\theta \partial_\theta + 2(r-M)\bar{\Delta} \partial_r\Bigg\}f_{\ell-2,m}^\nz(r,\theta)\nonumber\\
&+ \mu^2 (f_{\ell,m}^\nz + f_{\ell+2, m}^\nz + f_{\ell-2,m}^\nz)(r,\theta)= \frac{27\omega_\ell^2M^2\zeta\chi^2}{r_{\p}^{6} \bar{\Delta}^2}\left(M - \frac{r_{\p}}{2}\right)\Big[M \hat{A}(r_{\p}) \cos^2\theta + \hat{B}(r_{\p})\Big]P_\ell^m(\cos\theta)G_{\ell,m}^{0,\nz}(r)\nonumber\\
&+\frac{\omega_\ell^2 M^2\zeta\chi^2}{150528 \bar{\Delta}^2r_{\p}^{7}}\Big[\hat{C}(r_{\p})\cos^2\theta + \hat{D}(r_{\p})\Big]\left(\frac{r - r_{\p}}{r_{\p} - r_{\m}}\right)(r_{\p} - r_{\m}),P_\ell^m(\cos\theta)G_{\ell,m}^{0,\nz}(r).
\label{kgnear}
\end{align}
\end{widetext}

As in the far zone, we see that we have terms on the right-hand side of Eq.~\eqref{kgnear} proportional to $P_\ell^m(\cos\theta)$ and $\cos^2\theta P_\ell^m(\cos\theta).$
We perform the same procedure for the near zone in the previous section. Using the recurrence relation, we find the following coefficients:
\ba 
\tilde{A} = \hat{B}(r_{\p}) + \hat{A}(r_{\p})\frac{(-\ell+2\ell(1+\ell) - 2m^2)}{-3 + 4\ell(\ell+1)},  \\
\tilde{B} = \hat{A}(r_{\p})\frac{(\ell-1+m)(\ell+m)}{-1 + 4\ell^2}, \\
\tilde{C} = \hat{A}(r_{\p})\frac{(1+\ell-m)(2+\ell-m)}{3 + 4\ell(2+\ell)} , \\
\bar{A} = \hat{D}(r_{\p}) + \hat{C}(r_{\p})\frac{(-\ell+2\ell(1+\ell) - 2m^2)}{-3 + 4\ell(\ell+1)},  \\
\bar{B} = \hat{C}(r_{\p})\frac{(\ell-1+m)(\ell+m)}{-1 + 4\ell^2}, \\
\bar{C} = \hat{C}(r_{\p})\frac{(1+\ell-m)(2+\ell-m)}{3 + 4\ell(2+\ell)} .
\ea 
$\tilde{B}$, $\bar{B}$, $\tilde{C}$, and $\bar{C}$ are the analogous coefficients to $\tilde{A}$ and $\bar{A}$ for the $l-2$ and $l+2$ solutions, respectively. 

Lastly, we present the details of the integrands $\mathcal{J}_1$ and $\mathcal{J}_2$ in Eq.~\eqref{dCSNearLargez}. We have: 
\begin{widetext}
\begin{align}
\mathcal{J}_1 &= {}_2F_1\Big([-\ell, \ell+1], [1+ 2iP], -z\Big){}_2F_1\Big([-\ell-2iP, \ell+1-2iP], [1+ 2iP], -z\Big)\Big[z(r_{\p} - r_{\m}) + r_{\p}\Big]^2\nonumber \\ &\times \frac{(\Omega_1^\nz \tilde{A}_{\ell,m} + z\Omega_2^\nz \bar{A}_{\ell,m})}{D_{\nz}}, \\
\mathcal{J}_2 &= \frac{{}_2F_1\Big([-\ell, \ell+1], [1+ 2iP], -z\Big)^2 [z(z+1)]^{iP}z^{iP}(z+1)^{-iP}\Big[z(r_{\p} - r_{\m}) + r_{\p}\Big]^2(\Omega_1^\nz \tilde{A}_{\ell,m} + z\Omega_2^\nz \bar{A}_{\ell,m})}{D_{\nz}}.
\end{align}
\end{widetext}
where the denominator $D_\nz$ is 
\begin{widetext}
\begin{align}
D_{\nz}&=8z^2(z+1)^3\Bigg\{iPz\left[P^{2} + \frac{1}{4}(\ell^2 + \ell + 1) + \frac{1}{8}(\ell^2+\ell)\right]{}_2F_1\Big([-\ell, \ell+1], [1+ 2iP], -z\Big)\nonumber \\
&\times {}_2F_1\Big([-\ell-2iP + 1, \ell+2 -2i P], [2- 2iP], -z\Big)\nonumber \\
&- \left[ -\frac{z(\ell+1)(iP - 1/2)\ell}{4}{}_2F_1\Big([-\ell+1, \ell+2], [2+ 2iP], -z\Big) + iP\left(P^{2} + \frac{1}{4}\right) {}_2F_1\Big([-\ell, \ell+1], [1 + 2iP]\Big) \right]\nonumber \\
&\times {}_2F_1\Big([-\ell-2iP, \ell+1-2iP],[-2iP + 1].  -z\Big)\Bigg\}.
\end{align} 
\end{widetext}
\section{Details of Asymptotic Matching}
\label{matchingdetails}
Here we give further detail about the expansions presented in Section \ref{dCS} C.
\label{asymptdetails}
\subsection{Far Zone}
We will begin from Eq.~\eqref{farsoldCS}. First, note that the Whittaker functions can be written in terms of ordinary and confluent hypergeometric functions as follows:
\begin{align}
W(\nu, \ell +1/2, x) &\approx e^{-x/2}x^\ell U(\ell + 1 -\nu, 2\ell + 2, x) \nonumber \\
M( \nu, \ell +1/2, x)  &\approx  e^{-x/2}x^\ell {}_2F_1(\ell + 1 -\nu, 2\ell +2, x). 
\end{align} 
We then want the small-x expansions of these expressions. Let us call the small-x expansion of the confluent hypergeometric $\tilde{U}$ and note that the ordinary ${}_2F_1$ is approximately one in the small-x limit. The relevant expansions in U are 
\begin{align}
    \tilde{U}(\ell + 1 - \nu, 2\ell + 2, x) \approx \frac{\Gamma(-1-2\ell)}{\Gamma(-\ell -\nu)} + x^{-1-2\ell}\frac{\Gamma(2\ell + 1)}{\Gamma(1 + \ell - \nu)}, \\
     \tilde{U}(\ell - \nu, 2\ell + 2, x) \approx \frac{\Gamma(-1-2\ell)}{\Gamma(-1-\ell -\nu)} + x^{-1-2\ell}\frac{\Gamma(2\ell + 1)}{\Gamma(\ell - \nu)}
\end{align}
Then, the integrals become:\\

\begin{widetext}
\begin{align}
    \mathcal{I}_1(x) &\approx \frac{\tilde{U}(\ell+1 - \nu, 2\ell + 2, x)}{\tilde{U}(\ell +1 - \nu, 2\ell + 2, x)(1 + \ell + \nu) +\tilde{U}(\ell - \nu, 2\ell + 2, x) }\nonumber \\
    &= \frac{1}{1 + 2\ell}  + x^{2\ell+1}\frac{\Gamma(-1-2\ell)\Gamma(1+\ell - \nu)}{\Gamma(2 + 2\ell)\Gamma(-\ell-\nu)}, 
\end{align}
\end{widetext}
and
\begin{widetext}
\begin{align}
    \mathcal{I}_2 &\approx \frac{\tilde{U}(\ell+1 - \nu, 2\ell + 2, x)^2}{\tilde{U}(\ell +1 - \nu, 2\ell + 2, x)(1 + \ell + \nu) +\tilde{U}(\ell - \nu, 2\ell + 2, x) }\nonumber \\
    &= \frac{2\gamma(-1-2\ell)}{(1+2\ell)\Gamma(-\ell -\nu)} + x^{-2\ell -1}\frac{\Gamma(2\ell + 1)}{(1 + 2\ell)\Gamma(1 + \ell - \nu)}+ x^{2\ell +1} \frac{\Gamma(-1-2\ell)^2\Gamma(1 + \ell -\nu)}{\Gamma(2 + 2\ell)\Gamma(-\ell-\nu)^2}. 
\end{align}
\end{widetext}
We then perform the integrals and simplify the full expression to obtain
\begin{align}
    g_{\ell}^p \approx \frac{(1 + \ell)\Gamma(-2\ell-2)}{\ell\Gamma(-\ell-\nu)} x^{\ell-1} + \frac{\ell\Gamma(2\ell)}{(1 + \ell)\Gamma(1 + \ell -\nu)} x^{-\ell-2}.
\end{align}
Lastly, we define $\nu = n+1+\ell + \delta\nu$ and expand in small $\delta\nu$ to obtain Eq.~\eqref{dCSFarSmallx}.
\subsection{Near Zone}
Now consider the large r expansion of the near zone solution. From Eq.~\eqref{soldCSNear}, we note that the integrands $\mathcal{J}_1$ and $\mathcal{J}_2$ each have a common overall factor of complicated hypergeometric functions each multiplied by a different hypergeometric. We take the common factor, expand in large z and simplify. Then, expand each of the leftover hypergeometric functions as 
\begin{widetext}
\begin{align}
    {}_2F_1([-\ell, \ell+1], [1+2iP], -z) &\approx\frac{\Gamma(1 + 2\ell)\Gamma(1 + 2iP)}{\Gamma(1+\ell)\Gamma(1 + \ell + 2iP)}z^{\ell} + \frac{\Gamma(-1-2\ell)\Gamma(1+2iP)}{\Gamma(-\ell)\Gamma(-\ell+2iP)}z^{-\ell-1}, \\
     {}_2F_1([-\ell-2iP, \ell+1-2iP], [1-2iP], -z) &\approx\frac{\Gamma(1 + 2\ell)\Gamma(1 - 2iP)}{\Gamma(1+\ell)\Gamma(1 + \ell - 2iP)}z^{\ell} + \frac{\Gamma(-1-2\ell)\Gamma(1-2iP)}{\Gamma(-\ell)\Gamma(-\ell-2iP)}z^{-\ell-1} .
\end{align}
\end{widetext}
Then, performing the integrals, collecting terms and simplifying yields Eq.~\eqref{dCSNearLargez}.

\subsection{Matching}
Using well known properties of the $\Gamma$ function, we now provide the details of simplifying the resulting expression from the matching condition. From the matching, we obtain 
\begin{widetext}
\begin{align}
   \delta\nu_\ell \approx \frac{2^{1-2 \ell} (-1)^{1-n} [k(r_\p - r_\m)]^{2 \ell+1} \sin (\pi  \ell) \Gamma (-2 \ell-1)^2 \Gamma (\ell+2 i P+1) }{\Gamma \left(\ell+\frac{1}{2}\right)^2 \Gamma (n+1) \Gamma (-2 \ell-n-1) \Gamma (2 i P-\ell)}.
   \label{Eq:deltanuunsimp}
\end{align}
\end{widetext}
We have the relations:
\begin{align}
\Gamma(-z-n) &= (-1)^{n+1}\frac{\Gamma(-z)\Gamma(1+z)}{\Gamma(1+n-z)},\\
\frac{\pi}{\sin(\pi z) } &= \Gamma(1 - z)\Gamma(z), \\
\Gamma(n + 1/2)  &= \frac{(2n)! \sqrt{\pi}}{4^n n!}.
\end{align}
We also have the following:
\ba 
\Gamma(1 + \ell + 2i P) = \Gamma(2iP)\prod_{j=0}^{\ell}(j+ 2iP),  \\
\Gamma(-\ell+2iP)  = \Gamma(2iP)\prod_{j=1}^{\ell}(-j+2iP)^{-1},  \\
\ea 
which leads to
\be 
\frac{\Gamma(1 + \ell + 2 i P)}{\Gamma(-\ell+2iP)} = 2iP (-1)^\ell \prod_{j=1}^\ell (j^2 + 4P^2),
\ee 
Putting all of this together and simplifying Eq.~\eqref{Eq:deltanuunsimp} , we obtain Eq.~\eqref{deltanufinal}.

\section{Full d'Alembertian Operators}
\label{BoxFull}
Here, for completeness we show the full expression for $\Box_{\dCS}$ for the general Kerr + dCS metric.

Consider the metric
\be 
g_{\mu\nu} = g_{\mu\nu}^{K} + g_{\mu\nu}^{\dCS}\left[\mathcal{O}(\zeta\chi^2)\right].
\ee
We have that 
\begin{widetext}
\be 
\Box f(t,r,\theta,\phi) = \Big\{\Box_{K} + \Box_{\dCS}(\mathcal{O}\left[\zeta\chi)\right] \nonumber\\
+ \Box_{\dCS}\left[\mathcal{O}(\zeta\chi^2)\right]\Big\}f(t,r,\theta,\phi). 
\ee
\end{widetext} 
$\Box_\K$ is just the usual expression for the Kerr d'Alembertian, and we then have at $\mathcal{O}(\zeta\chi)$ 
\begin{widetext}
\be 
\Box_{\dCS}\left[\mathcal{O}(\zeta\chi)\right]f(t,r,\theta,\phi) = \nonumber \\
{\frac {{M}^{5} \left( 189\,{M}^{2}+120\,Mr+70\,{r}^{2} \right) 
 \zeta\,\chi}{56\,{r}^{6}\Delta \left( r
 \right) }}  {\frac {\partial ^{2}f \left( t,r,
\theta,\phi \right)}{\partial t\partial \phi}} .
\ee 
\end{widetext}
At $\mathcal{O}(\zeta\chi^2)$, we have the following: 
\begin{widetext}
\begin{align}
\Box_{\dCS}\left[\mathcal{O}(\zeta\chi)\right]f(t,r,\theta,\phi) &= \frac{27M^3\zeta\chi^2}{2r^{12}\Delta(r)^2\sin^2\theta}\Bigg[\frac{-27\sin^2\theta r\Delta(r)^2}{8} \Bigg(-\frac{40\Delta(r)}{9}\left(\cos^2\theta - \frac{1}{3}\right)A(r) + Mr^2 B(r)\Bigg) \frac{\partial^2 f}{\partial r^2} \nonumber \\
&+2\sin^2\theta r^6 \Bigg(\frac{\sin^2\theta M^3\Delta(r) }{8}C(r) + \left(M - \frac{r}{2}\right)\cos^2\theta D(r) + E(r)\Bigg) \frac{\partial^2 f}{dt^2}\nonumber \\
&- 2\sin^2\theta r \Delta(r)^2 \left(\cos^2\theta - \frac{1}{3}\right)F(r)\frac{\partial^2 f }{\partial\theta^2}\nonumber \\
&-2r\Delta(r)\Bigg(G(r)\cos^2\theta \Delta(r) + H(r) \Delta(r) - \frac{\sin^2\theta r^3 M^3}{8}I(r)\Bigg)\frac{\partial^2 f}{\partial \phi^2}\nonumber \\
&- \sin^2\theta \Delta(r) \Bigg(-15 r\left(\cos^2\theta - \frac{1}{3}\right)J(r) \Delta(r) + \left(M - \frac{r}{2}\right)r^2\Big(K(r)\cos^2\theta + L(r)\Big) \frac{d\Delta(r)}{dr}\nonumber \\
&+ 8\Delta(r) \Big(M(r)\Delta(r)\cos^2\theta + N(r)\Delta(r) + rO(r)\cos^2\theta + r P(r)\Big)\Bigg)\frac{\partial f}{\partial r}\nonumber \\
&+ 2\sin\theta \cos\theta r \Delta(r)\Bigg(-\frac{17M\cos^2\theta\Delta(r)}{2}Q(r) + R(r) \Delta(r) - \left(M - \frac{r}{2}\right)\sin^2\theta r S(r)\Bigg)\frac{\partial f}{\partial \theta},
\label{boxfarfull}
\end{align}
\end{widetext}
where the coefficients are given by 
\begin{widetext}
\begin{align}
A(r) &= {M}^{6}+{\frac {5623\,{M}^{5}r}{30240}}+{\frac {379\,{M}^{4}{r}^{2}}{
6048}}-{\frac {2579\,{M}^{3}{r}^{3}}{42336}}-{\frac {125\,{M}^{2}{r}^{
4}}{15876}}-{\frac {1459\,M{r}^{5}}{362880}}-{\frac {67\,{r}^{6}}{
40320}},
\\
B(r) &={M}^{5}-{\frac {523\,{M}^{4}r}{10206}}-{\frac {55\,{M}^{3}{r}^{2}}{972
}}-{\frac {805\,{M}^{2}{r}^{3}}{8748}}-{\frac {25\,M{r}^{4}}{5832}}-{
\frac {25\,{r}^{5}}{17496}},
 \\ 
 C(r) &={M}^{2}+{\frac {40\,Mr}{63}}+{\frac {10\,{r}^{2}}{27}} \\
 D(r) &={M}^{6}-{\frac {149\,{M}^{5}r}{2016}}+{\frac {445\,{M}^{4}{r}^{2}}{
2016}}+{\frac {235\,{M}^{3}{r}^{3}}{1568}}+{\frac {3725\,{M}^{2}{r}^{4
}}{21168}}+{\frac {67\,M{r}^{5}}{896}}+{\frac {67\,{r}^{6}}{2688}},
 \\
 E(r) &= -{\frac {{M}^{7}}{3}}+{\frac {275\,{M}^{6}r}{6048}}-{\frac {197\,{M}^{
5}{r}^{2}}{12096}}+{\frac {1745\,{M}^{4}{r}^{3}}{63504}}-{\frac {4145
\,{M}^{3}{r}^{4}}{254016}}+{\frac {1241\,{M}^{2}{r}^{5}}{254016}}+{
\frac {67\,M{r}^{6}}{16128}}+{\frac {67\,{r}^{7}}{16128}},
\\ 
F(r) &= {M}^{6}+{\frac {481\,{M}^{5}r}{2016}}+{\frac {5615\,{M}^{4}{r}^{2}}{
28224}}+{\frac {185\,{M}^{3}{r}^{3}}{6048}}+{\frac {4727\,{M}^{2}{r}^{
4}}{84672}}+{\frac {355\,M{r}^{5}}{12096}}+{\frac {67\,{r}^{6}}{5376}},\\
G(r) &= {M}^{6}+{\frac {229\,{M}^{5}r}{2016}}+{\frac {375\,{M}^{4}{r}^{2}}{
3136}}-{\frac {95\,{M}^{3}{r}^{3}}{6048}}+{\frac {4727\,{M}^{2}{r}^{4}
}{84672}}+{\frac {355\,M{r}^{5}}{12096}}+{\frac {67\,{r}^{6}}{5376}}
, \\
H(r) &= -{\frac {{M}^{6}}{3}}+{\frac {275\,{M}^{5}r}{6048}}+{\frac {1105\,{M}^
{4}{r}^{2}}{84672}}+{\frac {655\,{M}^{3}{r}^{3}}{18144}}-{\frac {4727
\,{M}^{2}{r}^{4}}{254016}}-{\frac {355\,M{r}^{5}}{36288}}-{\frac {67\,
{r}^{6}}{16128}},\\
I(r) &= {M}^{2}+{\frac {40\,Mr}{63}}+{\frac {10\,{r}^{2}}{27}}, \\
J(r) &= {M}^{6}+{\frac {5623\,{M}^{5}r}{30240}}+{\frac {379\,{M}^{4}{r}^{2}}{
6048}}-{\frac {2579\,{M}^{3}{r}^{3}}{42336}}-{\frac {125\,{M}^{2}{r}^{
4}}{15876}}-{\frac {1459\,M{r}^{5}}{362880}}-{\frac {67\,{r}^{6}}{
40320}}, \\
K(r) &={M}^{6}-{\frac {149\,{M}^{5}r}{2016}}+{\frac {445\,{M}^{4}{r}^{2}}{
2016}}+{\frac {235\,{M}^{3}{r}^{3}}{1568}}+{\frac {3725\,{M}^{2}{r}^{4
}}{21168}}+{\frac {67\,M{r}^{5}}{896}}+{\frac {67\,{r}^{6}}{2688}} ,\\
L(r) &= -{\frac {{M}^{6}}{3}}+{\frac {9473\,{M}^{5}r}{6048}}+{\frac {4115\,{M}
^{4}{r}^{2}}{6048}}+{\frac {17285\,{M}^{3}{r}^{3}}{63504}}-{\frac {
2255\,{M}^{2}{r}^{4}}{63504}}-{\frac {1459\,M{r}^{5}}{72576}}-{\frac {
67\,{r}^{6}}{8064}}, \\
M(r) &= {M}^{7}-{\frac {1157\,{M}^{6}r}{2304}}+{\frac {1039\,{M}^{5}{r}^{2}}{
5376}}+{\frac {5575\,{M}^{4}{r}^{3}}{225792}}+{\frac {8555\,{M}^{3}{r}
^{4}}{169344}}-{\frac {2237\,{M}^{2}{r}^{5}}{451584}}-{\frac {67\,M{r}
^{6}}{21504}}-{\frac {67\,{r}^{7}}{43008}},\\
N(r) &= -{\frac {67\,{r}^{6}}{43008}}-{\frac {6035\,{M}^{5}r}{12096}}-{\frac {
1381\,M{r}^{5}}{290304}}-{\frac {1975\,{M}^{4}{r}^{2}}{10752}}-{\frac 
{11135\,{M}^{2}{r}^{4}}{1016064}}+{\frac {45265\,{M}^{3}{r}^{3}}{
338688}}-{\frac {57\,{M}^{6}}{16}},\\
O(r) &= {M}^{7}-{\frac {1157\,{M}^{6}r}{2304}}+{\frac {1039\,{M}^{5}{r}^{2}}{
5376}}+{\frac {5575\,{M}^{4}{r}^{3}}{225792}}+{\frac {8555\,{M}^{3}{r}
^{4}}{169344}}-{\frac {2237\,{M}^{2}{r}^{5}}{451584}}-{\frac {67\,M{r}
^{6}}{21504}}-{\frac {67\,{r}^{7}}{43008}},\\
P(r) &=-{\frac {9931\,{M}^{6}r}{6912}}+{\frac {283\,{M}^{5}{r}^{2}}{5376}}+{
\frac {156175\,{M}^{4}{r}^{3}}{2032128}}+{\frac {8825\,{M}^{3}{r}^{4}
}{127008}}+{\frac {6157\,{M}^{2}{r}^{5}}{1354752}}+{\frac {953\,M{r}^{
6}}{580608}}+{\frac {67\,{r}^{7}}{129024}}-{\frac {{M}^{7}}{3}},\\
Q(r)&={M}^{5}+{\frac {139\,{M}^{4}r}{672}}+{\frac {220\,{M}^{3}{r}^{2}}{2499
}}-{\frac {3575\,{M}^{2}{r}^{3}}{79968}}-{\frac {13\,M{r}^{4}}{34272}}
-{\frac {13\,{r}^{5}}{137088}},\\
R(r) &= -{\frac {67\,{r}^{6}}{8064}}+{\frac {19343\,{M}^{5}r}{12096}}-{\frac {
2957\,M{r}^{5}}{145152}}+{\frac {26065\,{M}^{4}{r}^{2}}{42336}}-{
\frac {10273\,{M}^{2}{r}^{4}}{254016}}-{\frac {101705\,{M}^{3}{r}^{3}
}{254016}}+{\frac {47\,{M}^{6}}{6}},\\
S(r)&={M}^{6}-{\frac {149\,{M}^{5}r}{2016}}+{\frac {445\,{M}^{4}{r}^{2}}{
2016}}+{\frac {235\,{M}^{3}{r}^{3}}{1568}}+{\frac {3725\,{M}^{2}{r}^{4
}}{21168}}+{\frac {67\,M{r}^{5}}{896}}+{\frac {67\,{r}^{6}}{2688}}. 
\end{align}
\end{widetext}

\normalem
\bibliographystyle{apsrev}
\bibliography{master}

\end{document}